\documentclass[11pt, a4paper]{article}

\usepackage{jheppub}

\usepackage{mathrsfs}
\usepackage[T1]{fontenc}
\usepackage{mathpazo}
\usepackage{setspace}
\usepackage{amsfonts}
\usepackage{amssymb}
\usepackage{amsmath}
\usepackage{epsfig}
\usepackage{latexsym}
\usepackage{color}
\usepackage{graphicx}
\usepackage{nicefrac}
\usepackage[latin1]{inputenc}
\usepackage{pstricks}
\usepackage{slashed}
\usepackage{multirow}
\usepackage{hyperref}

\def\be{\begin{equation}}
\def\ee{\end{equation}}
\def\ba{\begin{eqnarray}}
\def\ea{\end{eqnarray}}

\def\del{\partial}

\def\k{\kappa}

\def\g{\gamma}
\def\G{\Gamma}

\def\D{\Delta}
\def\e{\epsilon}

\def\th{\theta}

\def\m{\mu}
\def\n{\nu}

\def\l{\lambda}
\def\L{\Lambda}
\def\s{\sigma}

\def\qq{\qquad}

\def\IR{\relax{\rm I\kern-.18em R}}

\def\inv{^{\raise.0ex\hbox{${\scriptscriptstyle -}$}\kern-.05em 1}}

\def \z { {\bar z} }

\allowdisplaybreaks[1]

\title{G-structures and Flavouring non-Abelian T-duality\footnote{MAD-TH-13-04}}

\author[a,b]{Alejandro Barranco,}

\author[c,d]{J\'er\^ome Gaillard,}

\author[b]{Niall T. Macpherson,}

\author[b,f]{Carlos N\'u\~nez,}

\author[g]{and Daniel C. Thompson}

\affiliation[a]{Institute of Cosmos Sciences and ECM Department, Facultat de Fisica, Universitat de Barcelona, 

 Av. Diagonal 647, 08028 Barcelona, Spain\\}

\affiliation[b]{Department of Physics, Swansea University, Singleton Park, Swansea, SA2 8PP, UK \\}

\affiliation[c]{Department of Physics, University of Wisconsin, Madison, WI 53706, USA\\}

\affiliation[d]{Institute for Advanced Study, Hong Kong University of Science and Technology, Hong Kong \\}

\affiliation[f]{CP3 -Origins and DIAS, University of 
Southern Denmark, Odense, Denmark \\}

\affiliation[g]{Theoretische Natuurkunde, Vrije Universiteit Brussel, and 

The International Solvay Institutes\\

Pleinlaan 2, B-1050, Brussels, Belgium}

 \emailAdd{alejandro@ecm.ub.edu}

 \emailAdd{jgaillard@wisc.edu}

  \emailAdd{pymacpherson@swansea.ac.uk}

    \emailAdd{c.nunez@swansea.ac.uk}

\emailAdd{dthompson@tena4.vub.ac.be}

\abstract{We examine non-Abelian T-duality on backgrounds preserving ${\cal N}=1$ supersymmetry showing how backgrounds with $SU(3)$-structure are mapped to backgrounds with $SU(2)$-structure. We provide the transformation rules for the corresponding pure spinors.  We use these techniques to construct new flavoured solutions of type IIA supergravity.}


\def\beq{\begin{equation}}

\def\eeq{\end{equation}}

\def\bea{\begin{eqnarray}}

\def\eea{\end{eqnarrat}}

\def\z{\zeta}

\def\h{\eta}

\def\e{\epsilon}

\begin{document}

\maketitle

\flushbottom

\section{Introduction}

Although the idea of generalising 
T-duality to non-Abelian isometry groups has rather old roots \cite{delaossa:1992vc}, it is only recently that it has been studied as full solution-generating symmetry of supergravity \cite{Sfetsos:2010uq,Lozano:2011kb,Itsios:2012dc,Lozano:2012au,Jeong:2013jfc,Itsios:2012zv,Itsios:2013wd}.  The recent work of Itsios et al. \cite{Itsios:2012zv,Itsios:2013wd} considered the application of this duality transformation in IIB supergravity backgrounds preserving ${\cal N}=1$ supersymmetry.    For instance applying an $SU(2)$ non-Abelian T-duality to the internal space of the Klebanov--Witten background ($AdS^5 \times T^{1,1}$) results in a solution of type IIA which retains the $AdS_5$ factor and has a lift to M-theory which corresponds to the geometries obtained in \cite{Bah:2012dg} from wrapping M5 branes on an $S^2$.   In \cite{Itsios:2013wd} similar dualisations were applied to non-conformal geometries (Klebanov--Tsetylin, Klebanov--Strassler and wrapped D5 models) resulting in a new class of smooth solutions of massive type IIA supergravity.  The field theory interpretation of these massive IIA solutions is, as yet, undetermined. However an analysis of the gravity solution indicates they retain rich RG dynamics displaying signatures of Seiberg duality, domain walls and confinement in the IR.

A common feature of the geometries obtained in  \cite{Itsios:2013wd} is that they retain four dimensional Poincar\'e invariance and it was argued that they should also retain ${\cal N}=1$ supersymmetry.  The conditions for a solution of type II supergravity to possess these symmetries can be very elegantly stated using the language of G-structures \cite{Grana:2004bg,Grana:2005sn}.   The existence of a single four-dimensional conserved spinor implies that on the six-dimensional internal manifold $M$ we have two spinors $\eta^1$ and $\eta^2$.   If these spinors are proportional, the structure group 
of $TM$, the tangent bundle of $M$,  is reduced to $SU(3)$ and can be characterised by an invariant real two-form $J$ and complex three-form $\Omega$ with $J\wedge \Omega = 0$ and $i\Omega \wedge \bar\Omega= \frac{4}{3} J^3$.  If on the other hand the two spinors are nowhere parallel they each define a separate $SU(3)$-structure and together equip $M$ with an $SU(2)$-structure consisting of a complex nowhere-vanishing vector field $v+ i w$, a real two-form $j$ and a complex two-form $\omega$.

These conditions can also be restated using the language of generalised complex geometry in which we consider the bundle $TM \oplus T^\ast M$.  The algebraic conditions of supersymmetry imply that there exist two pure spinors $\Psi_\pm = \eta_+^{1} \otimes \eta_\pm^{2\dag} $.   Using the Clifford map these pure spinors can be described as a formal  sum of forms, for instance in the case of $SU(3)$-structure we identify $\Psi_+ = e^{-i J}$ and $\Psi_- = \Omega$.   The differential conditions of supersymmetry can  be succinctly expressed in this language  (as closure conditions for the annihilator space of these pure spinors under the H-twisted Courant bracket) and  are schematically given by
\beq
	 d_H  \Psi_1 =  0 \ , \qquad d_H \Psi_2 = F_{RR} \ ,   
\eeq
where $d_H = d + H \wedge$, $F_{RR}$ denotes the RR fields and $\Psi_{1,2}$ are related to the pure spinors $\Psi_\pm$ depending on the type of supergravity in question.

This approach also makes clear the transformation rules under T-duality; these pure spinors essentially transform in the same way as Ramond fields.   Indeed, in the case where $M$ is  Calabi-Yau, mirror symmetry serves to interchange the pure spinors $e^{-i J} \leftrightarrow \Omega$.  The extension of this, {\it \`a la} Strominger, Yau and Zaslow \cite{Strominger:1996it},  
to $SU(3)$ structure compactifications has been developed in \cite{Gurrieri:2002wz,Fidanza:2003zi} and the case of general $SU(3)\times SU(3)$ structure considered in \cite{Grana:2006hr}.

The first purpose of this note is to study the effects of 
non-Abelian T-duality on these G-structures and thereby to give 
credence to the conjecture made in   \cite{Itsios:2013wd} that  
in general 
the result of the dualisation will be to take an 
$SU(3)$-structure background to one with $SU(2)$-structure.  
A heuristic reason for this can be found 
by looking at the abelian case following \cite{Grana:2008yw}.    After T-duality, left and right movers couple to different set of frame fields for the same geometry, call them $\hat{e}_+^i$ and $\hat{e}_-^i$.  In the simplest case we can understand this T-duality as a reflection on right movers so that in directions dualised $  \hat{e}_+^i = -  \hat{e}_-^i$.  The  $J$ and $\Omega$ of the starting $SU(3)$-structure give rise, after dualisation, to a $\hat{J}$ and $\hat{\Omega}$ which may be expressed in terms of either the left or right moving frame fields giving a corresponding $\hat{J}_\pm$ and $\hat\Omega_\pm$.  Suppose that the expression for $\hat{J}$ is 
\beq
 	\hat{J}_\pm = \hat{e}_\pm^1 \wedge \hat{e}_\pm^2 +  \hat{e}_\pm^3 \wedge \hat{e}_\pm^4+ \hat{e}_\pm^5 \wedge \hat{e}_\pm^6 \ . 
\eeq
Consider the case where the dualised directions are $1$ and $2$, then $\hat{J}_+  =\hat{J}_-$ and in this case the T-dual also has $SU(3)$-structure.  Now consider the dualisation of two directions that are not paired by the complex structure, say $1$ and $3$, in this case  $\hat{J}_+  \neq \hat{J}_-$ and type changing has occurred; the $SU(3)$-structure gives rise to a T-dual $SU(2)$-structure after T-dualisation.   Since the non-Abelian T-dualisations performed in  \cite{Itsios:2013wd} involve three directions they cannot respect the paring of the complex structure and so we anticipate them to be  type changing.   One goal of this paper is to make this reasoning precise and to provide explicit examples where the T-dual $SU(2)$-structure can be obtained.

The second part of this paper concerns a topic which at first sight might seem rather disconnected from the above discussion namely the application of non-Abelian T-duality in the construction of new `flavoured' solutions of supergravity.  The string dual view on the addition  of fundamental matter to the field theories has a rich history. Starting from the study of the `quenched' dynamics of fundamental fields,  equivalent to the addition of probe branes in the string backgrounds to the case in which flavour branes (sources) backreact and change the original geometries, various technical problems have been resolved. 
For reviews see \cite{Erdmenger:2007cm}, \cite{Nunez:2010sf}.

In the case of backgrounds preserving some amount of SUSY, the first technical point to be addressed is to find SUSY embeddings for these sources or flavour branes. The embeddings were initially found  solving differential equations associated with the kappa-symmetry matrix. A more refined and efficient way of expressing the same conditions relies on G-structures and calibration forms. Indeed, the findings of papers like \cite{Casero:2006pt} among many others can be thought as examples of the generic formalism developed in \cite{Martucci:2005ht},   \cite{Koerber:2007hd} and more  explicitly layed-out in \cite{Gaillard:2008wt}, \cite{Gaillard:2010qg}.

A generic feature about these solutions encoding the dynamics of $N_f$  fields transforming in the fundamental representation of the $SU(N_c)$ gauge group is that the string backgrounds  should in principle represent sources localised on those  SUSY-preserving submanifolds. The complications associated with the non-linear and coupled partial differential equations this problem requires, lead to the consideration of `smeared' sources---the field theoretical effect of such simplification is the explicit breaking of $SU(N_f)\to U(1)^{N_f}$. The SUSY-preserving way of implementing this smearing is also described by the G-structures  classifying the original (unflavoured) background, see \cite{Gaillard:2008wt}, \cite{Gaillard:2010qg} for details.

Hence, there is a rich interplay between G-structures and the dynamics of SUSY sources in Supergravity. This is one of the themes of this work.  Using the results established in the first part of the paper we will be able to construct the non-Abelian T-dual of a flavoured background.

We hope it is clear from the discussion above, that the main goal of this paper 
is to {\it geometrise} part of the information of the works \cite{Itsios:2012zv,Itsios:2013wd} .
The idea being that once in a geometric context, the physical analysis (to be done in the future)
will become more clear and systematic.  On the other hand, we emphasise the underlying motivation: the 'utility' of non-Abelian T-duality
is to produce backgrounds (hard obtain by an educated guess) that being smooth, they {\it define}
a dual QFT. So, understanding the geometric side of the non-Abelian T-duality will help characterise a set of new strongly coupled field theories.

The structure  of this paper is as follows: in section 2 we present some of the salient details of non-Abelian T-duality. In section 3 we provide some more details on $SU(3)$ and $SU(2)$-structures and their transformation rules under non-Abelian T-duality. 
In section 4 
we look at examples 
of the T-dual of the un-flavoured Klebanov-Witten model 
studied in \cite{Itsios:2012zv,Itsios:2013wd} and explicitly construct 
its $SU(2)$-structure. In section 5 we present 
the flavoured Klebanov-Witten model and its T-dual.

\section{Non-Abelian T-duality}

In this section we present some useful overview of non-Abelian T-duality,  
a comprehensive treatment may be found in \cite{Itsios:2013wd}.

The three-step Buscher procedure of gauging a $U(1)$ isometry, 
enforcing a flat connection for the corresponding gauge field with 
a Lagrange multiplier, and integrating out these Lagrange 
multipliers provides a powerful way to construct a T-dual $\sigma$-model. 
This approach can be readily generalised to the case of non-Abelian 
isometries and provides a putative non-Abelian T-duality transformation.    
Unlike its Abelian counter part, this non-Abelian T-duality 
typically destroys the isometries dualised (though they can 
be recovered as non-local symmetries of the string $\sigma$-model \cite{Curtright:1994be}).     
Due to global complications, it is thought that this non-Abelian 
dualisation is not a full symmetry of string (genus) perturbation 
theory however it remains valid as a solution-generating 
symmetry of supergravity. 
In this regard its status is rather similar to fermionic T-duality 
\cite{Berkovits:2008ic},
which has proven to be very useful in the 
context of the AdS-CFT correspondence in providing 
an explanation of the scattering amplitude/Wilson 
loop connection at strong coupling \cite{Alday:2007hr}.

Let us first consider a bosonic string $\sigma$-model in a NS background. We will assume that this background admits some isometry group $G$ and that background fields can be expressed in terms of  left-invariant Maurer--Cartan forms, $L^i= -i Tr(g^{-1} d g)$, for this group. That is to say the target space metric has a decomposition 
\beq
\label{gstart}
	ds^2 = G_{\mu \nu }(x) dx^\mu dx^\nu + 2 G_{\mu i}(x) L^i +  g_{ij}(x) L^i  L^j \ , 
\eeq
with corresponding
expressions for the NS two-form $B$ and dilaton $\Phi$.  
The  non-linear  $\sigma$-model is 
\beq
	S = \int d^2 \sigma \Big(
Q_{\m\n} \del_+ x^\m \del_-x^\n+  
Q_{\m i} \del_+ x^\m L_-^i + Q_{i\m} L_+^i  
\del_- x^\m + E_{ij} L_+^i L_-^j \Big) \ ,
\eeq
where
\be
Q_{\m\n} = G_{\m\n} + B_{\m \n} \ ,\quad Q_{\m i} = G_{\m i } + B_{\m i } \ , \quad  Q_{ i\m} = G_{ i\m } + B_{ i\m } \ , \quad E_{ij} = g_{ij} +  b_{ij}\ ,
\ee
and $L^i_\pm$ are the left-invariant forms pulled back to the world sheet.  
To obtain the dual $\sigma$-model one first gauges the isometry  by making the replacement
\be
\del_\pm  g \to D_\pm g = \del_\pm g - A_\pm g \ ,
\ee
in the Maurer--Cartan forms. 
Also, the addition of 
a Lagrange multiplier term $-i {\rm Tr}(v F_{+-})$  enforces 
a flat connection.

After integrating this Lagrange multiplier term by parts, 
one can solve for the gauge fields to obtain the T-dual model. 
Finally, we must gauge fix the redundancy 
by, for example,  setting $g= \mathbb{1}$
\footnote{More general gauge fixing choices are allowed and will 
in fact be exploited in this paper. 
For details of these we refer the reader to \cite{Itsios:2013wd}.  
In this section we assume the 
gauge fixing choice of  $g= \mathbb{1}$.}.

We obtain the Lagrangian,
\be
\tilde S  = \int d^2 \sigma \Big( Q_{\m\n}\del_+ x^\m \del_- x^\n + (\del_+ v_i + \del_+ x^\m Q_{\m i})( E_{ij} + f_{ij}{}^k v_k)^{-1} (\del_- v_j - Q_{j\m} \del_- x^\m ) \Big)\ ,
\label{tdulal}
\ee
from which the T-dual metric and $B$-field can be ascertained.  As with Abelian T-duality the dilaton receives a shift from performing the above manipulations in a path integral given by 
\be
	\hat\Phi(x, v)  = \Phi(x) - \frac{1}{2} \ln(\det M) 
\ee 
where we have defined $M_{ij} = E_{ij} + f_{ij}{}^k v_k$ which will play a prominent role in what follows.

Using the equations of motion, one can ascertain the following transformation rules for the world-sheet derivatives  
\be
\begin{aligned}
\label{worldsheettrans}
	& L^i_+ = -(M^{-1})_{ji} \left(\del_+ v_j + Q_{\m j}\del_+ x^\m \right)\ , \\
	& L_-^i = M^{-1}_{ij} (\del_- v_j - Q_{j\m} \del_- x^\m)\ , \\
	& \del_\pm x^\m ={\rm invariant}\ .
\end{aligned}
\ee
These relations provide a classical canonical 
equivalence between the two T-dual $\sigma$-models \cite{Curtright:1994be,Lozano:1995jx}.
 
The consequence of this is that left and right movers couple to different sets of vielbeins for the T-dual geometry. Suppose that we define frame fields for the initial metric \eqref{gstart} by
\be
 ds^2 = \eta_{AB} e^A e^B + \sum_{i = 1}^{\dim G} \delta_{ab} e^a e^b \ , \quad e^A = e^A_\mu  dx^\mu \ , \quad e^a =\kappa_i^a L^i + \lambda^a_\mu dx^\mu \ . 
\ee
Then by making use of the transformation rules \eqref{worldsheettrans} one finds that after T-dualisation left and right movers couple to the vielbeins
\be
\label{eq:Tdualframes}
\begin{aligned}
	&\hat{e}^a_+  =   -\kappa M^{-T} \big(dv + Q^T dx \big) + \l dx  \ , \quad \hat{e}^A_+ = e^A \\ 
	&\hat{e}^a_-  =  \kappa M^{-1} \big(dv - Q\, dx \big) + \l dx \ , \quad \hat{e}^A_- = e^A \ ,
 \end{aligned}
\ee
in which $M^{{-T}}$ is the inverse transpose of the matrix $M$ defined above.
Both these frame fields define the T-dual target space metric obtained from  \eqref{tdulal} given by
\be
\widehat{ds}^2 = \eta_{AB} e^A e^B + \sum_{i = 1}^{\dim G} \delta_{ab} \hat{e}_+^a \hat{e}_+^b  =  \eta_{AB} e^A e^B + \sum_{i = 1}^{\dim G} \delta_{ab} \hat{e}_-^a \hat{e}_-^b \ . 
\ee
Since these frame fields define the same metric they must be related by a Lorentz transformation and indeed
\be
\hat e_+ = \L \hat e_-\ , \quad \L = -\kappa M^{-T} M \kappa^{-1} \ .
\ee
We note that $\det \Lambda = (-1)^{\dim G}$, this will have the 
consequence that the dualisation of an odd-dimensional isometry 
group maps between type IIA and IIB theories whereas that 
of an even-dimensional group preserves the chirality.   
This Lorentz transformation induces an action on spinors defined 
by the invariance property of gamma matrices
\footnote{Unfortunately, the existing notation in the literature means 
we have the same symbol $\Omega$ for the 
spinorial transformation matrix and for the  
$SU(3)$-structure three-form. We trust the reader will infer from the context which is meant.}; 
\be
\Omega^{-1}  \Gamma^a  \Omega =  \Lambda^a{}_b \Gamma^b\ .
\label{spino1}
\ee

We are particularly interested in performing this duality in supergravity backgrounds of relevance to the AdS/CFT correspondence which are typically supported by RR fluxes.   Then one ought to, in principle, reconsider the above derivation in a formalism suitable of including RR fluxes. In the case of Abelian and Fermonic T-duality this has explicitly been done in the 
pure spinor approach  \cite{Benichou:2008it,Sfetsos:2010xa} and a simple extrapolation of these results to this non-Abelian context leads to the following conclusion which can also be motivated from the considerations of \cite{Hassan:1999bv}.   The dual RR fluxes are obtained by right multiplication by the above matrix $\Omega$ on the RR bispinor (this can be viewed equivalently as a Clifford multiplication on the RR polyform/pure spinor).  Explicitly,  the T-dual fluxes are given by \cite{Sfetsos:2010uq}:
\def\FF{{\bf F}}
\def\CC{{\bf C}}
\be
e^{\hat{\Phi}} \hat{\slashed{\FF} } =   e^\Phi \slashed{\FF} \cdot \Omega^{-1} \ , 
\label{ppom}
\ee
where the RR polyforms are defined by  
\be
{\rm IIB}:\  {\FF }  =  \sum_{n=0}^4  F_{2n+1}\ ,
\qq
{\rm IIA}:\ { \FF }  =  \sum_{n=0}^5 F_{2n}\ , 
\ee
and the slashed notation in equation \eqref{ppom} indicates that we have converted these polyforms to bispinors by contraction with gamma matrices.  Here we are working in the democratic formalism in which all ranks of fluxes are considered as independent and Hodge duality implemented by hand afterwards\footnote{See the appendices for details of the conventions used.}.

For many applications knowledge of the transformation 
laws for the gauge-invariant field strengths is sufficient.  
However, in some applications we will also be interested 
on how the RR potentials 
themselves transform.  We define potentials as  
\be
{\rm IIB}:\  {\CC }  =  \sum_{n=0}^4 C_{2n}\ ,
\qq
{\rm IIA}:\  { \CC }  =  \sum_{n=0}^4 C_{2n+ 1}\ , 
\ee
related to the field strengths by 
\be
{\rm IIB}:\  {\FF }  = ( d-H \wedge)\CC  \ .
\qq
{\rm IIA}:\  {\FF }  = ( d-H \wedge)\CC + m e^{B} \ , 
\ee
in which $m$ is the Romans mass parameter of type IIA. Actually we will need to be a bit more general than this when we consider the addition of sources, see appendix \ref{Sec: source conventions}.

We propose that the potentials so defined have a straightforward transformation rule: 
 \be
e^{\hat{\Phi}} \hat{\slashed{\CC} } =   e^\Phi \slashed{\CC} \cdot \Omega^{-1} \ . 
\label{cpom}
\ee
We should comment briefly about a subtlety; the potentials 
in the equation above
have to be chosen in such a way that the 
T-duality can be readily performed. 
In other words, for the transformation rule to be as above, the potentials $C_p$ should have a vanishing Lie derivative along the Killing vectors of the isometry dualised.  
 A less judicious choice of potentials would require composing the above transformation law with an appropriate gauge transformation that first brings the potential into the desired form (this is well explained in \cite{Grana:2008yw} for the NS two-form potential which need not have a vanishing Lie derivative under the isometry dualised but instead obey ${\cal L}_k B = d \xi$).  
 
 Although we have not shown that \eqref{cpom} implies \eqref{ppom} in all generality, we find that it does indeed generate the correct transformation in the case at hand.  The essential step in a general proof would be to show that the Clifford multiplication implied by the spinor contraction  in  \eqref{cpom} commutes with the action of the twisted differential $d_H$.  One may be confident that this is true in all generality since this is indeed the case with Abelian T-duality \cite{Grana:2008yw} and we shall see that in a certain basis the transformation rules do become very similar to the Abelian case.

We end this section by remarking the status of supersymmetry under 
non-Abelian T-duality.  Supersymmetry need not be preserved by T-duality (Abelian or not).\footnote{In principle, supersymmetry can even be enhanced by T-duality but given that non-Abelian T-duality destroys isometry this seems rather unlikely in this case.}  Whether (and how much) supersymmetry is preserved depends on how the Killing vectors about which we dualise act on the supersymmetry.  The action of a vector on a spinor, which is only well defined when the vector is Killing, is given by \cite{Kosmann}
\be\label{eq:Kosmann}
 {\cal L}_k \epsilon = k^\mu D_\mu \epsilon  + \frac{1}{4} \nabla_\mu k_\nu \gamma^{\mu \nu} \epsilon \ . 
\ee
If, when acting on the Killing spinor of the initial geometry,  this vanishes automatically for all the Killing vectors that generate the action of $G$ then we anticipate supersymmetry to be preserved in its entirety.  If on the other hand this vanishes only for some projected subset of Killing spinors then we expect only a corresponding projected amount of supersymmetry to be preserved in the T-dual.\footnote{In \cite{Itsios:2012dc} this was confirmed to be true in general for a large class of backgrounds.}   In this paper we consider the case of ${\cal N}= 1$ supersymmetry which is invariant under the above action of $G$ so that the non-Abelian duality should preserve supersymmetry.  Suppose we start  with ten-dimensional  MW Killing spinors $\epsilon^1$ and $\epsilon^2$, then the Killing spinors in the T-dual will be given by 
\be
\label{Tdualspinors}
\hat{\epsilon}^1 = \epsilon^1 \ , \quad \hat{\epsilon}^2  = \Omega\cdot\epsilon^2 \ . 
\ee

\section{G-structures and their transformations}

We now give a brief summary of the important details concerning G-structures. We follow the conventions of \cite{Martucci:2005ht} except where indicated otherwise.  We consider ten-dimensional backgrounds consisting of a warped product of four-dimensional Minkowski space and a six-dimensional internal manifold $M$: 
\be
ds^2_{10} = e^{2A} ds^2_{1,3}  + ds^2(M) \ . 
\ee
Since we require ${\cal N }=1$ 
supersymmetry there should exist a single four-dimensional 
conserved spinor.  The ten-dimensional MW spinors of type II supergravity are decomposed as
\be
\begin{aligned}
	& \e^1 = \zeta_+ \otimes \eta^1_+ +   \zeta_- \otimes \eta^1_-\ , \\
	& \e^2 = \zeta_+ \otimes \eta^2_\mp  +   \zeta_- \otimes \eta^2_\pm \ , 	   
\end{aligned}
\ee 
where the upper sign in $\e^2$ corresponds to IIA and the lower to IIB --
here $\pm$ denotes both four and six-dimensional chiralities and we choose a basis such that $(\eta_+)^\ast = \eta_-$.   From the internal spinors we define two $Cliff(6,6)$ pure spinors (or polyforms):
\be
\Psi_\pm = \eta^1_+ \otimes (\eta_\pm^2)^\dag \ .
\ee
We define the norms of the internal spinors $|| \eta^1 ||^2 = |a|^2 $ and  $|| \eta^2 ||^2 = |b|^2 $.   
The dilatino and gravitino equations can be recast 
succinctly, for the type IIA case, as
\be \label{eq:purespinoreqs}
\begin{aligned}
	  e^{-2 A + \Phi}  (d+ H\wedge) \! \left[  e^{2 A - \Phi}  \Psi_-  
\right]  \!&= 
dA \wedge \bar \Psi_-  
+ \frac{e^\Phi}{16} \left[ \left(|a|^2 \!- \! |b|^2\right)F_{IIA,-} \!
+ i  \left(|a|^2 \! + \! |b|^2\right) \! \star_6 F_{IIA,+}\!  \right] \!, \\ 
	  (d+ H\wedge) \! \left[  e^{2 A - \Phi}  \Psi_+  \right]  \!&=  0 \ .
\end{aligned}
\ee 
The RR fluxes entering on the right-hand side of this equation 
are defined for the type IIA case as
\be 
F_{IIA,-}= F_0-F_2 + F_4 - F_6,\;\;\; F_{IIA,+}= F_0+ F_2+ F_4 + F_6.
\ee
Similar expressions hold in the case of type IIB after exchanging $\Psi_+ \leftrightarrow \Psi_-$ and $F_{IIA} \leftrightarrow F_{IIB}$, see \cite{Grana:2004bg}
 and \cite{Martucci:2005ht} for details.

Two important extreme cases are when the internal spinors 
are always parallel (corresponding to $SU(3)$-structure) and when they are 
nowhere parallel (that corresponds to 
$SU(2)$-structure).  In the first case there is a single 
spinor of unit norm such that $\eta^1_+ = a \eta_+ $, $\eta^2_+ = b \eta_+$.  The spinor bilinears then define a two-form and a complex three-form with components
\be 
J_{mn}  = - \frac{i}{|a|^2} \eta_+^{1\dag} \gamma_{mn} \eta_+^1 \ , \qquad \Omega_{mnp}  = - \frac{i}{a^2} \eta_-^{1\dag} \gamma_{mnp} \eta_+^1 \ . 
\ee
These are normalised such that $J^3 = \frac{3i}{4} \Omega \wedge \bar \Omega$ and obey $J\wedge \Omega = 0 $.   The corresponding pure spinors are
\be\label{Eq:SU3 pure spinor}
SU(3)~structure: \quad \Psi_+  = \frac{ ab^* }{8} e^{-i J} \ , \quad  \Psi_-  = - \frac{i ab }{8} \Omega \ . 
\ee

In the second case when the spinors are nowhere parallel we have a non-vanish complex vector field defined by $\eta^1_+  =  a\eta_+ $, $\eta^2_+ = b (v^i + i w^i)\gamma_i \eta_-$.  In this case one can show that the corresponding pure spinors have the form 
\be\label{eq:SU2purespinor}
SU(2)~structure: \quad \Psi_+  = \frac{ ab^*}{8} e^{-i  v\wedge w } \wedge \omega \ , \quad  \Psi_-  =  \frac{ab}{8} e^{-ij} \wedge (v + i w)  \ . 
\ee 
We can express the forms $v$, $w$, $\omega$ and $j$ directly in terms of the spinors (see for example \cite{Mariotti:2007ym}):
\beq
	\begin{aligned}
		v_m - i w_m &= -\frac{1}{a b} \eta_-^{2\dag} \gamma_m \eta_+^1 \,, \\
		\omega_{mn} &= \frac{i}{a b^*} \eta_+^{2\dag} \gamma_{mn} \eta_+^1 \,, \\
		j_{mn} &= \frac{i}{2|b|^2} \eta_+^{2\dag} \gamma_{mn} \eta_+^2 - \frac{i}{2|a|^2} \eta_+^{1\dag} \gamma_{mn} \eta_+^1 \,.
	\end{aligned}
\eeq

To ascertain the non-Abelian T-dual of these structures one can work explicitly with the T-dual Killing spinors defined in equation \eqref{Tdualspinors} and construct from first principles the pure-spinors $\Psi_\pm$ defined above.  Alternatively, for the spinor-phobic one can circumvent this by using the following transformation rules on the polyforms 
\beq\label{eq:CliffordMult}
	\slashed{\Psi}^{SU(2)}_{+} =   i  \,  \slashed{\Psi}^{SU(3)}_{-}\Omega^{-1}  \,, \quad \slashed{\Psi}^{SU(2)}_{-} =   \slashed{\Psi}^{SU(3)}_{+}\Omega^{-1}  \ .
\eeq 
The D-brane generalised calibrations follows from this as shown in appendix \ref{su2structuresxxx}.

 Let us just remark at this stage that the condition of supersymmetry being preserved as detailed 
in equation \eqref{eq:Kosmann} simply translates (using the Liebniz derivation property obeyed the Lorentz-Lie derivative \cite{Kosmann}) into the invariance of the pure-spinors under the regular Lie derivative acting on forms:
\be
 {\cal L}_k \epsilon = 0 \Rightarrow  {\cal L}_k \Psi_\pm = 0 \ . 
\ee
For the case of the abelian T-duality one can show that this criteria does indeed ensure that supersymmetry is preserved after T-duality  \cite{Grana:2008yw}.  The essence of the proof is that up to terms proportional to this Lie derivative,  the twisted differential $d_H$ commutes with the Clifford multiplication rule (c.f. equation \eqref{eq:CliffordMult})  used to extract the T-dual pure spinors.  Using this,
 one can infer that supersymmetry is preserved by the dualisation.  Although we have not verified the details the situation here appears to be exactly analogous, indeed as we shall shortly see one can find a basis in which the non-Abelian T-duality essentially mimics the Abelian case. 

In the following sections, we will consider 
two examples that will make clear various points discussed above.
The first case-study will be the non-Abelian T-dual of
the Klebanov-Witten system as presented in 
\cite{Itsios:2012zv,Itsios:2013wd}. 
We will explicitly show the $SU(2)$-structure of the solution
(and hence its SUSY preservation).
We will then consider the background obtained by 
adding fundamental fields (quarks) to the 
Klebanov-Witten field theory \cite{Benini:2006hh} 
(conversely, we will consider the addition 
of source-branes to the  Klebanov-Witten background).
With the essential 
help of the $SU(2)$-structure formalism 

\section{Example 1: Unflavoured Klebanov--Witten and its T-Dual}

In this section we shall examine the T-dual of the Klebanov--Witten geometry and explicitly demonstrate its $SU(2)$-structure.

The theory living on D3 branes placed at the tip of the conifold was studied by Klebanov and Witten in \cite{Klebanov:1998hh}.  The gauge theory describing the low-energy dynamics of the branes is an ${\cal N}=1$ superconformal field theory with product gauge group $SU(N) \times SU(N)$.  It can be described by a two-node quiver and has two sets of bi-fundamental matter fields $A_i$ in the $(N,\bar{N})$ representation of the gauge group and $B^m$ in the $(\bar{N}, N)$. The indices $i$ and $m$ correspond to two sets of $SU(2)$ global symmetries.  The super potential for the matter fields is given by
\be
	W = \frac{\lambda}{2} \e^{ij} \e_{mn} Tr\left(A_i B^m A_j B^n\right) \ . 
\ee

This gauge theory is dual to string theory on $AdS^5 \times T^{(1,1)}$ with $N$ units of RR flux supporting the geometry:
\be
	\begin{aligned}
		& ds^2 = \frac{r^2}{L^2} dy^2_{1,3} + \frac{L^2}{r^2} dr^2 + L^2 ds^2(T^{(1,1)}) \ ,  \\
		& F_{(5)} = \frac{4}{g_s L} \left( \rm{vol}(AdS_5 ) - L^5 \rm{vol}(T^{(1,1)})   \right) \ . 	
	\end{aligned}
\ee
We will work with the following frame fields for this geometry 
\be
	\begin{aligned}
		& e^{y^\mu} = \frac{r}{L} dy^\mu  \quad (\mu = 0\dots 3) \ , \quad e^r = \frac{L}{r}dr  \ , \quad e^{\varphi} = \lambda_1 \sin\theta d \varphi  \ , \quad e^\th =  \lambda_1   d \theta  \ , \\
		& e^{1} = \lambda_1 \sigma_1 \ , \quad e^2 = \lambda_1 \sigma_2 \ , \quad e^3 = \lambda \left( \sigma_3 + \cos\theta d\varphi \right)  \ , 
	\end{aligned}
\ee
in which $\lambda_1^2 = \frac{L^2}{6}$ and $\lambda^2 = \frac{L^2}{9}$ and  we have introduced $SU(2)$ left-invariant one-forms parametrised by Euler angles:
\begin{equation}
\label{eq:leftinvforms}
	\sigma_1=(-\sin\psi d\tilde{\theta}+\cos\psi\sin\tilde{\theta} d\tilde{\varphi} ), \quad
	\sigma_2=(\cos\psi d\tilde{\theta}+\sin\psi\sin\tilde{\theta} d\tilde{\varphi} ), \quad
	\sigma_3=(\cos\tilde{\theta} d\tilde{\varphi}+d\psi).
\end{equation}
For reference we state the ten-dimensional spinors of KW in this basis given by
\beq
\label{eq:KWspinors}
	\begin{aligned}
		\e_1 = \sqrt{\frac{r}{L}} \Big( \z_+ \otimes \h_+ + \z_- \otimes \h_- \Big) \ , \quad \e_2 = \sqrt{\frac{r}{L}} \Big(i\, \z_+ \otimes \h_+ - i \, \z_- \otimes \h_- \Big) \ . 
	\end{aligned}
\eeq
The chiralities in these expressions are defined with respect to four and six-dimensional chirality matrices 
\beq
		\g_{(4)} = i \, \g^{y^0y^1y^2y^3}\ , \quad  \g_{(6)} = -i \, \g^{\varphi\theta123r} \ , 
\eeq
such that under the ten-dimensional 
chirality operator  $\G_{(10)} = \g_{(4)} \otimes \g_{(6)}$ 
both $\e_1$ and $\e_2$ are positive. 
In addition the spinor $\h_+$ is 
constant and normalised such that $\h_+^\dagger \h_+ = 1$.  
Supersymmetry imposes the following projections on the spinor
(as above $\eta_+= (\eta_-)^*$),
\beq 
\label{eq:KWprojections}
	\g^{r3} \h_+ = \g^{12} \h_+ = \g^{\varphi\theta} \h_+ = -i\,\h_+ \ .
\eeq
Using these expressions, we can determine the $SU(3)$-structure of KW in this basis to be
\beq
	\begin{aligned}
		J &= e^{\theta \varphi} - e^{12} + e^{3r} \ ,  \\
		\Omega &= (e^2 + i\, e^1) \wedge (e^{\theta} + i\, e^{\varphi}) \wedge (e^3 + i\, e^r) \ . 
	\end{aligned}
\label{eq:aguero}
\eeq
The non-Abelian T-dual of this geometry with respect to the $SU(2)$ global symmetry defined by the $\sigma_i$ was constructed in \cite{Itsios:2012zv,Itsios:2013wd}.  The result is an ${\cal N}=1$ supersymmetric solution of type IIA whose NS sector is given by\footnote{We have set $L=1$ which may be restored by appropriate rescalings.  Also in deriving these results the gauge fixing choice is different to that described in section 2 of this paper.  Details may be found in \cite{Itsios:2013wd}.  } 
\be
\label{KWTdual}
	\begin{aligned}
		d\hat s^2 = & ds_{\rm AdS_5}^2 + \l_1^2 (d\theta^2 + \sin^2 \theta d\varphi^2 )+ \frac{\l_1^2\l^2}{\D} x_1^2 \hat{\s}_3^2 \\
 		&  + \frac{1}{\D} \left( (x_1^2 + \l^2 \l_1^2 )dx_1^2 + (x_2^2 + \l_1^4) dx_2^2 + 2 x_1 x_2 dx_1 dx_2   \right) \ ,\\ 
		\hat B = & - \frac{\l^2}{\D} \left[x_1 x_2 dx_1  + (x_2^2 + \l_1^4) dx_2 \right]\wedge\hat{\s}_3  \ , \\ 
		e^{-2 \hat \Phi} = & \frac{8}{g_s^2}\ \D\ ,
	 \end{aligned}
\ee
where $ \hat{\s}_3 = d\psi + \cos\th d\varphi$ and
\beq
	\D \equiv  \l_1^2 x_1^2 + \l^2 (x_2^2 + \l_1^4 ) \ .
\label{KWTdualDelta}
\eeq

The metric evidently has an $SU(2) \times U(1)_\psi$ isometry and for a fixed value of $(x_1 , x_2)$ the remaining directions give a squashed three-sphere.  This geometry is supported by two and four-form RR fluxes which may be computed using equation \eqref{ppom} and whose explicit form can be found in  \cite{Itsios:2013wd}.  We remark in passing that the lift of this geometry to eleven dimensions has an interpretation in terms of recently discovered ${\cal N}=1$ SCFT's obtained from wrapping M5 branes on a Riemann surface (of genus zero in this case) \cite{Bah:2012dg}.

One can establish the left and right-moving T-dual frames for this geometry along the lines of equation \eqref{eq:Tdualframes}.  The frames in the $AdS$ direction are unaltered as are $e^\theta$ and $e^\varphi$.  In the directions dualised we find new frame fields $\hat{e}_\pm^i$ for $i = 1 \dots 3$.  
The plus and minus T-dual frames are related by a Lorentz transformation which, as described in section 2, induces a transformation on spinors given by \footnote{The careful reader will not confuse this matrix $\Omega$ and its inverse $\Omega^{-1}$ with the complex three-form defining an $SU(3)$-structure, that appears for example in equation \eqref{eq:aguero}.},
\beq 
\label{eq:Omega}
	\Omega= \frac{\G_{(10)}}{\sqrt{\D}} \Big( -\l_1^2 \l \G^{123} + \l_1 x_1 \cos \psi  \, \G^1+ \l_1 x_1 \sin \psi \,  \G^2 + \l x_2 \G^3 \Big) \ . 
\eeq
This defines the Killing spinors of the T-dual to be 
\beq
	\hat \e_1 = \e_1 \ ,  \qquad \qquad  \hat \e_2 = \Omega\cdot \e_2 \ . 
\eeq
Implementing the four-six decomposition one finds from \eqref{eq:KWspinors} using \eqref{eq:KWprojections} that 
\beq
	\begin{aligned}
		\hat \e_1 &= \sqrt{\frac{r}{L}} \Big( \z_+ \otimes \h_+ + \z_- \otimes \h_- \Big) \ , \\
		\hat \e_2 &= \sqrt{\frac{r}{L}} \Big( \z_+ \otimes \hat \h^2_- + \z_- \otimes \hat \h^2_+ \Big) \ ,
	\end{aligned}
\eeq
where
\beq
	\hat \h^2_- = -\frac{i}{\sqrt{\D}} 
\Big( \l_1^2 \l  \g^{r} + \l_1 x_1 \cos \psi \,   \g^1 + \l_1 x_1 
\sin \psi \,   \g^2 + \l x_2  \g^3 \Big) \h_+ \ , \;~~~ 
\hat{\eta}^2_+ = (\hat{\eta}^2_-)^*.
\eeq 
It is clear that in this basis, the T-dual Killing spinors depend not only on the radial coordinate but also on the T-dual coordinates $x_1, x_2$.   It is helpful to work in a different basis in which this new spinor can be expressed as simply as possible. In addition, we would like the new vielbein basis to preserve the geometric structure defined by $\h_+$, because $\epsilon_1$ is invariant under the non-Abelian T-duality. To do so we perform a rotation to a new basis $\tilde{e} = R \hat e$ (ordered as $r,  \varphi, \theta , 1 , 2 ,3$) with the rotation matrix 
\beq 
\label{eq:hattotilde}
	R = \frac{1}{\sqrt{1+\zeta.\zeta}}\left(
	\begin{array}{cccccc}
		1 & 0 & 0 & \zeta^1 & \zeta^2 & \zeta^3 \\
		0 & \sqrt{1+\zeta.\zeta} & 0 & 0 & 0 & 0 \\
		0 & 0 & \sqrt{1+\zeta.\zeta} & 0 & 0 & 0 \\
		-\zeta^1 & 0 & 0 & 1 & -\zeta^3 & \zeta^2 \\
		-\zeta^2 & 0 & 0 & \zeta^3 & 1 & -\zeta^1 \\
		-\zeta^3 & 0 & 0 & -\zeta^2 & \zeta^1 & 1 \\
	\end{array}
	\right)
\eeq
with,
\be
	\zeta^1=\frac{x_1 \cos \psi}{\l\l_1}\ , ~~~~~~\zeta^2=\frac{x_1 \sin \psi}{\l\l_1}\ , ~~~~~~\zeta^3=\frac{x_2}{\l_1^2} \ . 
\ee
Notice that these parameters are reflecting the structure of the spinor transformation matrix $\Omega$. 
The rotated vielbeins are given, in coordinate frame, by:
\beq
	\begin{aligned}
		\tilde{e}^r &= \frac{\lambda  \lambda _1^2dr -r (x_1dx_1 +x_2dx_2)}{r \sqrt{\Delta}} \ , & \tilde{e}^{\varphi} &= \lambda _1 \sin \theta \, d\varphi \ ,\\
		\tilde{e}^{1} &= \lambda _1 \frac{r \l (x_1 \sin \psi \, \hat{\s}_3 - \cos \psi \, dx_1) - x_1 \cos \psi \, dr}{r \sqrt{\Delta}} \ , &\tilde{e}^{\theta} &= \lambda _1 d\theta   \ , \\
		\tilde{e}^{2} &= -\lambda _1 \frac{r \l (x_1 \cos \psi \, \hat{\s}_3 + \sin \psi \, dx_1) + x_1 \sin \psi dr}{r \sqrt{\Delta}} \ ,  & \tilde{e}^{3} &= -\frac{\lambda  x_2 dr + \lambda _1^2 r \, dx_2}{r \sqrt{\Delta}} \ . 
	\end{aligned}
\eeq  
Then in this new basis (in which the gamma matrices are of course also rotated $\tilde \g = R  \g$),  we can easily show that
\beq
	\begin{aligned}
		\tilde \e_1 &= \sqrt{\frac{r}{L}} \Big( \z_+ \otimes \h_+ + \z_- \otimes \h_- \Big) \ , \\
		\tilde \e_2 &= \sqrt{\frac{r}{L}} \Big( \z_+ \otimes \tilde \h^2_- + \z_- \otimes \tilde \h^2_+ \Big) \ ,
	\end{aligned}
\eeq
with $\tilde{\eta}^2_+ = (\tilde{\eta}^2_-)^*$ and,
\beq
	\tilde \h^2_- = -i\, \tilde \g^r \h_+ \ . 
\eeq
Note that, as required for type IIA supergravity, the new spinors have opposite chirality. With this simple relation between $\tilde \h^2_-$ and $\h_+$, we clearly see that they are never parallel, hence we have an $SU(2)$-structure. Because we were careful about the definition of our new vielbein basis, the projections on $\eta_+$ are not modified, 
\beq 
	\tilde \g^{r3} \h_+ = \tilde \g^{12} \h_+ = \tilde \g^{\varphi\theta} \h_+ = -i\,\h_+ \ ,
\eeq
but the projections obeyed by $\tilde \h^2_-$ are different 
\beq 
	-\tilde \g^{r3} \tilde \h^2_- = \tilde \g^{12}\tilde \h^2_- = \tilde \g^{\varphi\theta} \tilde \h^2_- = -i\,\tilde \h^2_- \ . 
\eeq
The Killing spinors define two different $SU(3)$-structures
\beq
	\begin{aligned}
		J^1 &= \tilde e^{\theta \varphi} + \tilde e^{21} - \tilde e^{3r} \ , \\
		\Omega^1 &= (\tilde e^2 + i\, \tilde e^1) \wedge (\tilde e^{\theta} + i\, \tilde e^{\varphi}) \wedge (-\tilde e^3 + i\, \tilde e^r) \ , \\
		J^2 &= \tilde e^{\theta \varphi} + \tilde e^{21} + \tilde e^{3r} \ ,  \\
		\Omega^2 &= (\tilde e^2 + i\, \tilde e^1) \wedge (\tilde e^{\theta} + i\, \tilde e^{\varphi}) \wedge (-\tilde e^3 - i\, \tilde e^r) \ , 
	\end{aligned}
\eeq
whose intersection is the $SU(2)$-structure given by 
\beq\label{eq: SU2forms}
	\begin{aligned}
		v + i w &= -\tilde e^3 + i \tilde e^r  \ , \\ 
		j &= \tilde e^{\theta \varphi} + \tilde e^{21} \ ,  \\
		\omega &= (\tilde e^2 + i\, \tilde e^1) \wedge (\tilde e^{\theta} + i\, \tilde e^{\varphi}) \ . 
	\end{aligned}
\eeq
An explicit check shows that these do 
indeed satisfy the dilatino and gravitino equations 
that follow from equation \eqref{eq:purespinoreqs}.

Note that it makes sense to mix $e^r$ with $e^1$, $e^2$ and $e^3$ when performing the rotation \eqref{eq:hattotilde} because the geometric structure links $e^r$ and $e^3$ in the projection $\g^{r3} \h_+ = -i\,\h_+$. Actually the choice of this rotation appears clearer when considering that, because of the geometric structure, the transformation of the spinor $\epsilon_2$ can be written 
very easily as $\Omega \, \epsilon_2 = -\tilde \Gamma^r \epsilon_2$.
It is in this new basis that the transformation
closely resembles the  T-duality of the Abelian case.

\section{Example 2: Flavoured Klebanov Witten and its T-Dual.}\label{sectionkwflavored}

 An important step if one is to try and use the AdS/CFT paradigm to understand QCD-like dynamics 
is to incorporate fundamental flavours (quarks) into the gauge theory and corresponding gravity descriptions. A first step in this direction is to add a finite number $N_f$ of fundamental flavours which in the IIB set-up is typically 
achieved by the inclusion of a finite number of flavour D7 branes. This is the probe or quenched limit;  
the colour D3 branes generate the geometry but the flavour branes 
do not back-react  and only minimise their world-volume (DBI) action 
without deforming the geometry.   Remarkably one can even work beyond this 
quenched approximation by allowing a large number of flavour branes ($N_f \sim N_c$) 
in which case the D7 branes deform the geometry, see \cite{Nunez:2010sf}
for reviews.

In the case at hand we will consider adding $N_f$ D7 branes to the 
KW geometry in such a way that supersymmetry is preserved.  
We first describe the gauge theory engineered from the D3-D7 system in the conifold.  
We consider D7 branes parallel to the D3 stack in the Minkowski directions 
with the remaining four directions embedded holomorphically and non-compactly 
in the conifold.   The strings that run between the D7 and the D3 give rise to massless flavours. 
To avoid gauge anomalies
on the field-theory side of the description
and supergravity tadpoles on the string side of it, 
one must include two branches 
of D7 branes giving rise to fundamental chiral superfields for 
each gauge group ($q, \tilde{q}$ in the $(N,1)$ and $(\bar{N},1)$ and $Q, \tilde{Q}$ in the $(1, N)$ and $(1, \bar N)$).   The superpotential for this theory is given by \cite{Benini:2006hh},
\be
	W = \frac{\lambda}{2} \e^{ij} \e_{mn} Tr\left(A_i B^m A_j B^n\right)  +  h_1 \tilde q^a A_1 Q_a + h_2 \tilde{Q}^a B_1 q_a  \ . 
\ee
Notice that the $SU(2)$ global symmetries are explicitly broken by the embedding of the D7 branes - this symmetry
will be recovered 
by smearing the sources, when we go beyond the probe limit.   The addition of flavours implies that the theory loses conformality; a positive beta function is generated and 
  {\it a priori} one expects a Landau pole in the UV.

We now turn to the gravity description.  By considering the $\k$-symmetry projectors one can determine that the 
supersymmetric embeddings of D7 branes in the KW background to lie along two branches (the $y^\mu$ denote the Minkowski directions) 
\cite{Benini:2006hh},
\begin{equation}
	\xi=(y^{\mu},r,\psi,S^2),~~~~\tilde{\xi}=(y^{\mu},r,\psi,\tilde{S}^2),
\end{equation}
where $S^2$ and $\tilde{S}^2$ are the two-spheres parametrised 
by $\theta$, $\varphi$ and $\tilde{\theta}$, $\tilde{\varphi}$ respectively.  
To avoid the D7 charge tadpole we must include $N_f$ branes on both branches.  
One can write an action for the whole system consisting of supergravity together with DBI and WZ terms of the 
D7 branes (in string frame)  
 \be \label{eq:SDBIWZ}
 	\begin{aligned}
 		S_{DBI} &= -T_{D7} \sum_{Nf} \int_\xi d^8\sigma e^{-\Phi} 
\sqrt{|P[g]|}   -
T_{D7} \sum_{Nf} \int_{\tilde{\xi}} d^8\sigma e^{-\Phi} \sqrt{|P[g]|}  \ , \\
 		S_{WZ} &= T_7  \sum_{Nf} \int P[C_8] \ ,  
 	\end{aligned}
 \ee
where $P$ indicates the pull-back to the appropriate cycle, sometimes also denoted below as  
$g \bigg\lvert_{\xi}$ .
We do not activate the gauge field on the brane itself and since there 
is no NS two-form in this geometry the WZ term is simple.  Now we consider the case where the 
number of flavour branes goes to infinity in which case they can be smeared. 
In other words we consider that each stack is distributed homogeneously 
across the two-sphere it does not wrap.\footnote{This smearing procedure overcomes 
the bound on the number of D7 branes that comes from 
looking the deficit angle of the D7 solution so $N_f$ may 
indeed be taken large.}   
In a field-theory perspective the $U(N_f)$ flavour symmetries are broken to their maximal torus.  
The supergravity effect can be encoded by introducing a {\em smearing form}:
  \begin{equation}
	\Xi_2 = -\frac{N_f}{4\pi}\left(\sin\theta d\theta\wedge d\varphi+\sin\tilde{\theta}d\tilde{\theta}\wedge d\tilde{\varphi}\right) \ . 
\end{equation}
The smearing procedure essentially boils down to 
replacing the DBI and WZ contributions of equation \eqref{eq:SDBIWZ} with 
 \be \label{eq:SDBIWZsmeared}
 	\begin{aligned}
 		S_{DBI} &\rightarrow -T_{D7} \sum_{Nf} \int d^{10}x  
e^{-\Phi} \left( \sin \tilde\theta \sqrt{|P[g]|}   + \sin \theta \sqrt{|P[g]|} \right)  \ , \\
 		S_{WZ} &\rightarrow T_7  \sum_{Nf} \int \Xi_2 \wedge C_8 \ . 
 	\end{aligned}
 \ee
One consequence of this smearing is that the Bianchi identities are modified 
\begin{equation}
	dF_1 =\Xi_2 \ , \qquad  dF_5 =0 \ . 
\end{equation}
The D7 brane back-reaction is accommodated by the following ansatz (as above  we work in  string frame)
  \begin{equation}
	\begin{aligned}
		 ds^2& =\frac{e^{\frac{\Phi}{2}}}{\sqrt{h}}dy^2_{1,3}+e^{\frac{\Phi}{2}} \sqrt{h}\bigg(dr^2+\lambda^2_1e^{2g}(\sin^2\theta d\varphi^2+d\theta^2)+\lambda^2_2e^{2g}(\sigma_1^2+\sigma_2^2)+\lambda^2 e^{2f}(\sigma_3+\cos\theta d\varphi)^2\bigg)\ , \\
		F_1&=\frac{N_f}{4\pi}(\sigma_3+\cos\theta d\varphi)
		\ ,\qquad  F_5 =(1+\star)dt\wedge dx^1\wedge dx^2\wedge dx^3\wedge Kdr \ , 
	\end{aligned}
\end{equation}
where the warp factors $f$, $g$, $h$ and the dilaton $\Phi$ are functions 
of the radial variable 
$r$ and $\lambda^2_1=\lambda^2_2=1/6$, $\lambda^2=1/9$ 
and as a consequence of the Bianchi identities $Kh^2e^{4g+f}=27\pi N_c$.
\footnote{The unflavoured  Klebanov-Witten  can be recovered with the following:
\begin{equation*}
y^{\mu}\to\frac{1}{\sqrt{g_s}}y^{\mu}\ ,~~N_f=0\ ,~~h=\frac{L^4}{g_sr^4}\ ,
~~e^{2f}=e^{2g}=r^2\ ,~~K=\frac{4r^3g_s}{L^4}\ ,~~e^{\Phi}=g_s \ . 
\end{equation*}}  
The $\sigma_i$'s are $SU(2)$ left invariant one-forms defined in 
equation \eqref{eq:leftinvforms}.  A convenient basis of vielbeins is given by:
\beq
\begin{aligned}
e^{y^\mu}&=e^{\nicefrac{\Phi}{4}}h^{-\nicefrac{1}{4}}dy^\mu\ , & e^r&=e^{\nicefrac{\Phi}{4}}h^{\nicefrac{1}{4}}dr \ , \\
e^{\varphi}&=\lambda_1e^{g+\nicefrac{\Phi}{4}}h^{\nicefrac{1}{4}}\sin\theta d\varphi \ ,& 
e^{\theta}&=\lambda_1e^{g+\nicefrac{\Phi}{4}}h^{\nicefrac{1}{4}}d\theta \ , \\
e^{1}&=\lambda_1e^{g+\nicefrac{\Phi}{4}}h^{\nicefrac{1}{4}}\sigma_1 \ , & e^{2}&=\lambda_1e^{g+\nicefrac{\Phi}{4}}h^{\nicefrac{1}{4}}\sigma_2 \ ,\\
e^3&=\lambda h^{\nicefrac{1}{4}}e^{f+\nicefrac{\Phi}{4}}(\sigma_3+\cos\theta d\varphi)\ .
\end{aligned}
\eeq
Like the unflavoured version, this solution supports an $SU(3)$-structure: 
\begin{equation}
\begin{aligned}
J&=-\left(e^{r3}+e^{\varphi\theta}+e^{12}\right)=-\frac{4\pi\sqrt{h}}{3 N_f}e^{\frac{\Phi}{2}}\left(\frac{1}{2}e^{2g}\Xi_2+ e^f dr\wedge F_1\right),\\
\Omega&=(e^2+i e^1)\wedge(e^{\theta}+i e^{\varphi})\wedge(e^3+i e^{r}).
\end{aligned}
\end{equation}
With these and the structure conditions for $SU(3)$ it is possible to derive a set of first order BPS equations for the various functions introduced thus far:
\begin{align}
f'&=e^{-f}(3-2e^{2f-2g})-\frac{3N_f}{8\pi}e^{\Phi-f}\ , \quad  g'=e^{f-2g}\ , \label{eq:BPS}\\
h'&=-27\pi N_c e^{-f-4g}\ , \quad \Phi'=\frac{3N_f}{4\pi}e^{\Phi-f}\ . \nonumber
\end{align}
The RR potentials can be expressed in terms of  the $SU(3)$-structure forms as:
\begin{equation}
C_8= -\frac{1}{2}e^{-\Phi}\left(\frac{e^{\Phi}}{h}vol_4\right)\wedge J\wedge J \ , 
\qquad  C_4=e^{-\Phi}\left(\frac{e^{\Phi}}{h}vol_4\right)\ , \
\end{equation}
where $F_9=\star F_1$. The reason why we did not cancel both factors of the dilaton
is just for comparison with formulas below.

Finally for the brane embedding to be supersymmetric it must obey the calibration condition:
\begin{equation}
\sqrt{-g_{\xi}}d^8\xi=-\frac{1}{2}\left(\frac{e^{\Phi}}{h}vol_4\right)\wedge J\wedge J \bigg\lvert_{\xi};~~~~\sqrt{-g_{\tilde{\xi}}}d^8\tilde{\xi}=-\frac{1}{2}\left(\frac{e^{\Phi}}{h}vol_4\right)\wedge J\wedge J \bigg\lvert_{\tilde{\xi}};
\end{equation}
where $\hat{g}_{\xi}$ is the induced metric on $\xi$ whilst $\bigg\lvert_{\xi}$ indicates the pull back onto $\xi$, and similarly for $\tilde{\xi}$.
This allows the DBI and WZ actions of the smeared brane embedding to be expressed as:
\begin{equation}\label{eq:DBIcalibrated}
S_{DBI}=\frac{1}{2}\int_{\mathcal M_{10}}e^{-\Phi}\left(\frac{e^{\Phi}}{h}vol_4\right)\wedge J\wedge J\wedge\Xi_2\ ,\qquad S_{WZ}=\int_{\mathcal M_{10}} C_8\wedge \Xi_2\ , 
\end{equation}
from which it is immediate that $S_{DBI}+S_{WZ}=0$, as required by SUSY. As the sources are calibrated the dilaton equation of motion, 
Einstein's equations and the flux equation for $H$ are all satisfied once the Bianchi identities are imposed. This is proved for any $SU(3)\times SU(3)$-structure background in \cite{Koerber:2007hd}.

We will now find the non-Abelian T-dual of this system involving
metric, fluxes and sources. The interest of this problem is two-fold. On the one hand, 
it teaches us  the effect of the non-Abelian duality on the Born-Infeld-Wess-Zumino action. On the other hand, it will
tell us how to find the new smearing forms. 
Both these points give clues to
a generic procedure.

\subsection{The T-dual}

We perform the non-Abelian T-duality along the $SU(2)$ directions 
as before.   
To compactly display the results it is convenient to perform a supplementary rotation as detailed in equation (3.21) of \cite{Itsios:2013wd}.
We find the frame 
fields for the T-dual metric to be
\begin{align}
\hat{e}^{1}&=-\frac{\lambda_1}{\Delta}e^{g+\frac{\Phi}{4}}h^{\nicefrac{1}{4}}\left((\lambda_1^2\lambda^2he^{2f+2g+\Phi}+x_1^2)dx_1+x_1x_2(d x_2+\lambda^2\sqrt{h}e^{2f+\frac{\Phi}{2}}\hat{\sigma}_3)\right) \ , \nonumber \\
\hat{e}^{2}&=\frac{\lambda_1}{\Delta}e^{g+\frac{3}{4}\Phi }h^{\nicefrac{3}{4}}\left(\l^2 x_2e^{2f}dx_1-\l_1^2x_1e^{2g}(dx_2+\l^2\sqrt{h}e^{2f+\frac{\Phi}{2}}\hat{\sigma}_3)\right) \ , \label{eq:KWFlavTdualframes} \\
\hat{e}^{3}&=-\frac{\lambda }{\Delta}e^{f+\frac{\Phi}{4}}h^{\nicefrac{1}{4}}
\left(x_1x_2 dx_1+(\l_1^4he^{4g+\Phi}+x_2^2)
dx_2-\l_1^2\sqrt{h}x_1^2e^{2g+\frac{\Phi}{2}}\hat{\sigma}_3\right) \ . \nonumber
\end{align}
Where we recall $\hat{\sigma}_3=\cos\theta d \phi +d \psi$  and 
\begin{equation}
\Delta=\sqrt{h}e^{\frac{\Phi}{2}}\left(\l_1^4\l^2he^{2f+4g+\Phi}+\l_1^2x_1^2e^{2g}+\l^2x_2^2e^{2f}\right) \ . 
\end{equation}
The T-dual NS sector is then given by 
\be\label{eq:KWFlavTdualNSsector}
\begin{aligned}
	d\hat{s}^2 &= (e^{y_\mu})^2 
+(e^r)^2+(e^{\varphi})^2+(e^{\theta})^2+(\hat{e}^{1})^2+(\hat{e}^{2})^2+(\hat{e}^{3})^2 \ , \\
	\hat B & =\frac{\lambda e^{f-g} x_2  }{\lambda_1x_1}\hat{e}^{13}+\frac{\lambda\lambda_1  e^{f+g+\frac{\Phi }{2}} \sqrt{h}}{x_1} \hat{e}^{23}  \ ,  \\ 
	  H&=d \hat{B}\ ,  \\ 
	e^{-2\hat\Phi} &= 8\Delta e^{-2\Phi} \ . 
\end{aligned}	
\ee
This geometry is supported by RR fluxes, obtained using the 
general formula equation \eqref{ppom}, 
\be\label{eq:KWFlavTdualRRsector}
\begin{aligned}
	F_0&=\frac{N_f}{\sqrt{2}\pi}x_2 \ , \\
	F_2&=\frac{\l_1 e^{g-f-\frac{\Phi}{2}}}{\sqrt{2}\l\pi}
\left(4\pi \l_1\l^2Ke^{2f+g}h^{\nicefrac{3}{2}}e^{\varphi\theta}+\l\l_1N_fe^{f+g+\Phi}\sqrt{h}\hat{e}^{12}-x_1N_fe^{\frac{\Phi}{2}}\hat{e}^{13}\right) \ , \\
	F_4&=-2\sqrt{2} e^{-\Phi} h K e^{\varphi\theta}\wedge
\left(\l x_2e^{f}\hat{e}^{12}+\l_1 x_1 e^{g}\hat{e}^{23}\right) \ . 
\end{aligned}	
\ee
Although there is an $F_0$, it is possible that one should not regard this as 
a solution of 
Massive IIA -- the would-be mass parameter 
is neither constant nor quantised--- 
but rather, as we shall discuss, this should be thought 
of as a solution to type IIA in  the presence of D8 sources.  
Now since the original Bianchi identities were not satisfied 
(due to D7 source) one would not expect these 
new fluxes in equation (\ref{eq:KWFlavTdualRRsector})
to obey standard Bianchi identities after the non-Abelian T-duality.  
Indeed, one finds T-dual smearing forms enter the game  
\be\label{eq:KWFlavTdualBianchi}
\begin{aligned}
	dF_0&=\Xi_1 \ , \\
	dF_2-F_0 H&=\Xi_1\wedge B+\Xi_3 \ ,\\
	dF_4-H\wedge F_2&=\frac{1}{2}\Xi_1\wedge B\wedge B+B\wedge\Xi_3 \ .
\end{aligned}	
\ee
We find a rather nice result: the T-dual smearing forms can be calculated directly as 
\begin{align}
\Xi_1&=-\frac{N_fe^{-g-\frac{\Phi}{4}}}{\sqrt{2}\pi\l_1 h^{\nicefrac{1}{4}}} 
\left(x_1\hat{e}^{2}+\l\l_1 \sqrt{h} e^{f+g+\frac{\Phi}{2}}\hat{e}^3\right)=\frac{N_f}{\sqrt{2}\pi}dx_2 \ , \nonumber \\
\Xi_3&=\frac{N_fe^{-2g-\frac{\Phi}{4}}}{\pi h^{\nicefrac{1}{4}}}e^{\varphi\theta}\wedge\left(\sqrt{3}x_1e^g \hat{e}^1+\sqrt{2}x_2e^{f}\hat{e}^3\right) \label{smearingformsvvv}
 \\
&=\frac{N_f}{\sqrt{2}\pi}\sin\theta \left(x_1d\theta\wedge d\varphi\wedge dx_1 +x_2d\theta\wedge d\varphi\wedge d x_2\right) \ . \nonumber
\end{align}
These may be obtained equally using a transformation rule much like that of the RR fields,
 \begin{equation}
e^{\Phi}\slashed{\Xi}_2\Omega^{-1} =e^{\hat{\Phi}}\hat{\slashed{\Xi}}_B \ ,
\end{equation}
where $\hat{\Xi}_B=e^{B}\wedge(\Xi_1+\Xi_3)$.  
The active smearing forms indicate sources for both D6 and D8 branes.

\subsection{A nice subtlety.}

There is  a subtlety here. 
A naive reasoning would lead us to believe 
that when the non-Abelian T-dual is applied to
D7 sources, it will generate charge 
for D8, D6, D4 branes, whilst in equation (\ref{smearingformsvvv}) we only have
D8, D6 charges, since $\Xi_5$, 
the smearing form for D4 charges is absent in equation (\ref{eq:KWFlavTdualBianchi}). 
Below, we will solve
this apparent contradiction.

If we consider the Bianchi identity of the RR polyform 
\beq
dF-H\wedge F=\hat{\Xi}\wedge e^{B} \ , 
\eeq
it is clear that since the LHS of this equation is gauge invariant 
the RHS must also be. 
Throughout this note we have set to zero gauge fields on the world-volume however one should remember that they occur in conjunction with 
the NS two-form in the gauge-invariant combination 
${\cal F} = B + 2\pi \alpha' d A$.  Then
the most conservative view is that performing 
a gauge transformation on the NS
$B$-field simply activates appropriate 
compensating world-volume gauge field. 
There is however another point of view which is 
to keep the world-volume gauge
fields turned off and instead compensate for 
a $B$-field transformation with
an appropriate redefinition of the smearing form 
$\hat{\Xi}$.  This is best thought of not as a gauge transformation but rather as a mapping. In this picture
the transformation of the NS potential, $B\to B+\Delta B$, mediates a redistribution 
of source charge between the D4 and
D6 branes. The reason to prefer this second viewpoint
is that turning on a one-form gauge field on the brane
would break either the $SU(N_f)$ or the $U(1)^{N_f}$ symmetry.

To explain this second viewpoint,
we consider the transformation $B\to B'= B+\Delta B$. 
Such a transformation must be supplemented by a transformation 
of the smearing polyform $\hat{\Xi}\to \hat{\Xi}'$ so that the Bianchi 
identity of the RR polyform is unchanged. This requires that 
\beq
\hat{\Xi'}\wedge e^{B'}=\hat{\Xi}\wedge e^B \ .
\eeq
As an example, consider a 
transformation for which $\Xi_1\wedge \Delta B =0$. Then  we still have 
\beq
dF_0=\Xi_1,~~~dF_2-H F_0 =\Xi_3+B\wedge\Xi_1\ .
\eeq
The final Bianchi identity of the RR sector then becomes
\beq
dF_4-H\wedge F_2=\Xi_5+B\wedge \Xi_3+\frac{1}{2}B\wedge B \wedge \Xi_1 \ ,
\eeq
where $\Xi_5= \Delta B\wedge \Xi_3$. 
So we generate an explicit source for D4 branes under such a transformation. 
Clearly there are always source D8 branes but whether we 
have explicit source D6's or source D6 and D4's  is
a gauge-dependent statement. We 
do not believe it is possible to find a gauge in which we only have explicit D8 sources. 
This appears to be related to the fact that the original type IIB D7 brane 
embedding has two branches. This may seem rather mysterious, however one should understand 
that the total DBI and WZ actions of the source branes depend 
only on the sources through the gauge-invariant quantity  $\Xi\wedge e^{B}$. The higher potentials in the WZ action, $C_5$, $C_7$ and $C_9$, are gauge invariant as consequence of the $SU(2)$ SUSY conditions (see appendix \ref{su2structuresxxx} for details on this). 
So, it is only the `portion' of the sources that are viewed 
as being explicit rather than induced that changes, the equations of motion, the Bianchi identities and the total Maxwell charge are all invariants.

In summary, we advocated a picture in which gauge transformations mediate a 
redistribution of the source charge between the D4 and D6 branes.  
This could be thought of as an `inverse' of the
Myers effect.

To emphasize these points above, we can consider their Page charges \cite{Marolf:2000cb} defined as
\beq
\begin{aligned}
Q^{D6}_{page}&=\int_{\mathcal{M}_2}(F_2-F_0 B) \ ,\\
Q^{D4}_{page}&=\int_{\mathcal{M}_4}(F_4-B\wedge F_2+\frac{1}{2}F_0 B\wedge B) \ .\\
\end{aligned}
\eeq
The Maxwell charges are invariant under 
a shift in the $B$-field described above. 
While the shift of the Page charges is given by
\beq
\begin{split}
\Delta Q^{D6}_{page} &= \int_{\mathcal{M}_2} F_0 \Delta B \ ,\\
\Delta Q^{D4}_{page} &= \int_{\mathcal{M}_4}(-\Delta B\wedge( F_2- F_0 B) +\frac{1}{2}F_0\Delta B \wedge \Delta B) \ .
\end{split}
\eeq
As these these integrals are defined over compact manifolds these quantities are invariant for small gauge transformations. The integrands are exact so the integrals are zero. It is of course a generic feature of Page charges that they are only defined up to quantised shifts under large gauge transformations\footnote{Large gauge transformations are topological in nature and always induce quantised shifts.}. This is generally interpreted in the literature as a Seiberg duality in the dual gauge theory as in \cite{Benini:2006hh}.

\subsection{Potentials, $SU(2)$-structure and Calibration.}

We may use the formula for the T-dual RR potential in equation \eqref{cpom} to 
find the RR potentials. These are given in coordinate 
frame by (for alternative expressions  see below)
\beq
	\begin{aligned}
C_5 &=e^{-\hat{\Phi}}\bigg(\frac{e^{\Phi}}{h}vol_4\bigg)\wedge\bigg(\frac{\l\l_1^2e^{f+2g+\Phi}hdr-(x_1dx_1+x_2dx_2) }{\sqrt{\Delta}}\bigg) \ ,   \\ 
C_7 & = e^{-\hat{\Phi}}\bigg(\frac{e^{\Phi}}{h}vol_4\bigg)\wedge\bigg(
\frac{\l_1^2e^{2g+\Phi} h \sin\theta d\theta\wedge d\varphi\wedge
(\l e^f x_2dr+\l_1^2e^{2g}dx_2)}{\sqrt{\Delta}}+  \\
&~~~~~~~~~~~~ \frac{\l\l_1^2x_1e^{f+2g+\frac{3\Phi}{2}}h^{\frac{3}{2}}
(\l_1^2e^{2g}(x_1dr\wedge d x_2+\l e^{f}dx_1\wedge dx_2)-
\l^2e^{2f}x_2 dr\wedge dx_1)\wedge \hat{\s}_3}{\Delta^{3/2}}\bigg) \ , \\
C_9 & = e^{-\hat{\Phi}}\bigg(\frac{e^{\Phi}}{h}vol_4\bigg)\wedge
\bigg(\l\l_1^4e^{f+4g+\frac{3\Phi}{2}}x_1h^{\frac{3}{2}}\sin\theta d\theta\wedge d\varphi\wedge\hat{\s}_3\bigg)\wedge  \\
&~~~~~~~~~~~~\bigg(\frac{(h \lambda ^2 \lambda _1^2 e^{2 f+2 g+\Phi }+x_1^2)dr\wedge dx_1+x_1x_2 dr\wedge dx_2+\l e^fx_2dx_1\wedge dx_2}{\Delta^{3/2}}\bigg)  \ .
	\end{aligned}
\eeq
This background is again of $SU(2)$-structure where the 
defining forms $v+iw$, $j$, $\omega$ are the same as in the unflavoured case
-- see equations (\ref{eq: SU2forms}) -- 
the only difference being that the parameters entering the rotation matrix used in equation \eqref{eq:hattotilde} become 
\beq
\zeta^1=\frac{e^{-f-g-\frac{\Phi}{2}}x_1 \cos \psi}{\l\l_1\sqrt{h}};~~~~~\zeta^2=\frac{e^{-f-g-\frac{\Phi}{2}}x_1 \sin \psi}{\l\l_1\sqrt{h}};~~~~~\zeta^3=\frac{e^{-2g-\frac{\Phi}{2}}x_2}{\l_1^2\sqrt{h}} \ . 
\eeq 
This rotation leads to the following simple vielbeins for the dual geometry
\beq
	\begin{aligned}
\tilde{e}^r &= \frac{ h \lambda  \lambda _1^2 e^{f+2 g + 
\Phi}dr-
(x_1 dx_1+x_2 dx_2)}{\sqrt{\Delta }},~~~\tilde{e}^{\varphi} = h^{\frac{1}{4}} \lambda _1 
e^{g+\frac{\Phi }{4}} \sin \theta  d\varphi ,~~~\tilde{e}^{\theta} = h^{\frac{1}{4}} \lambda _1 e^{g+\frac{\Phi }{4}} d\theta \ , \\
\tilde{e}^{1} &= 
\sqrt{h} \lambda _1 e^{g+\Phi/2} \frac{-x_1 \cos \psi \ , 
dr - e^{f} \lambda 
(\cos \psi \, dx_1 - x_1 \sin \psi \, \hat{\s}_3) }{\sqrt{\Delta }} \ , \\
\tilde{e}^{2} &= -\sqrt{h} \lambda _1 e^{g+\Phi/2} \frac{x_1 \sin \psi \, dr + e^{f} \lambda (\sin \psi \, dx_1 + x_1 \cos \psi \, \hat{\s}_3) }{\sqrt{\Delta }} \ , \\
\tilde{e}^{3} &= -\sqrt{h} e^{\frac{\Phi}{2}}\frac{\lambda e^f x_2 dr + \lambda_1^2 e^{2 g}dx_2}{\sqrt{\Delta }} \ .  
\end{aligned}	
\eeq
This whole background is indeed a solution to the 
combined (massive)-IIA supergravity plus DBI plus WZ action (the details are
explicit in appendix \ref{Sec: source conventions}):
\beq
S=S_{\rm Massive\ IIA} +S_{DBI}+S_{WZ} \ .
\eeq
In the gauge in which the $B$-field is given by equation
(\ref{eq:KWFlavTdualNSsector}) 
and there are no explicit D4 sources, the appropriate WZ terms are given by
\begin{align}
S_{WZ}&=S^{D8}_{WZ}+S^{D6}_{WZ} \ , \nonumber \\
S^{D6}_{WZ}&=\int_{M_{10}}\!\!\bigg(C_7-B\wedge C_5\bigg)\wedge\Xi_3\ , \label{WZ1} \\
S^{D8}_{WZ}&=-\int_{M_{10}}\!\!\left(C_9-B\wedge C_7+\frac{1}{2}B\wedge B\wedge C_5\right)\wedge\Xi_1 \ , \nonumber
\end{align}
whilst the DBI action, expressed in terms of the D8 and D6 calibrations 
-- c.f. \eqref{eq:DBIcalibrated} -- is given by
\begin{equation}
\begin{split}
S_{DBI}&=S^{D8}_{DBI}+S^{D6}_{DBI},\\
S^{D6}_{DBI}&=-\int_{M_{10}}\!\!\!\! 
e^{-\hat{\Phi}}\left(\frac{e^{\Phi}}{h}vol_4\right)\wedge\bigg(v_1\wedge j_2-w_1\wedge B\bigg)\wedge\Xi_3, \\
S^{D8}_{DBI}&=-\int_{M_{10}}\!\!\!\! e^{-\hat{\Phi}}\left(\frac{e^{\Phi}}{h}vol_4\right)\wedge
\left(\frac{1}{2}w_1\wedge j_2\wedge j_2+v_1\wedge j_2\wedge B-\frac{1}{2}w_1\wedge B\wedge B\right)\wedge\Xi_1 .\\\
\label{WZ}
\end{split}
\end{equation}
Operating with the SU(2) structure we can recast the RR potentials as 
\beq\label{eq:CintermsofSU2}
\begin{split}
C_5 &= e^{-\hat{\Phi}}\big(\frac{e^{\Phi}}{h}vol_4\big)\wedge w_1 \ ,\\ 
C_7 & =e^{-\hat{\Phi}}\big(\frac{e^{\Phi}}{h}vol_4\big)\wedge j_2\wedge v_1  \ ,\\
C_9 & =-\frac{1}{2}e^{-\hat{\Phi}}\big(\frac{e^{\Phi}}{h}vol_4\big)\wedge j_2\wedge j_2 \wedge w_1 \ .
\end{split} 
\eeq
This makes it clear that on shell, as is required by sypersymmetry,  $S_{DBI} + S_{WZ} =  0 $.  This reflects the fact that the branes are calibrated, a fact that we now discuss in some detail. 
%
%

\subsection{Analysis of the dualised geometry}
One is often interested, particularly in the context of the AdS/CFT correspondence, in the possibility that D-branes may wrap certain submanifolds of the geometry in a way that preserves supersymmetry.  One approach to check whether a brane embedding is supersymmetric is to look carefully at the $\kappa$-symmetry projectors.  An alternative approach is the use of calibrations.  We recall that a calibration $\varpi$ is a closed $l$-form that bounds the volume of any oriented $l$-dimensional submanifold $\Sigma$ by
\beq
d^l \sigma \sqrt{\det g|_\Sigma} \geq \varpi |_\Sigma \ .
\eeq
A  submanifold is said to  be calibrated when  this bound is saturated and it follows that such a calibrated cycle will have the minimal volume within its homology class.   Of course in the geometries described above we have both NS and RR fluxes and this simple calibration is not enough to establish supersymmetric D-brane configurations.  For this one needs a {\em generalised calibration}, $\varpi$  which is a $d_H = d + H \wedge$ closed polyform  such that for any D-brane with world-volume field strength ${\cal F} =   B|_\Sigma + 2\pi \alpha' d A$  wrapping an internal cycle $\Sigma$, one has 
\be
{\cal E} \geq  \varpi |_\Sigma \wedge e^{{\cal F}} \ , 
\ee
where ${\cal E}$ is the energy density of the D-brane. When this bound is saturated the D-brane minimises its energy and is supersymmetric.  $SU(3) \times SU(3)$ backgrounds  admit a rich structure of supersymmetric cycles and the polyforms $\Psi_\pm$ (or rather the appropriate imaginary parts) serve as generalised calibrations as detailed by Martucci and Smyth in \cite{Martucci:2005ht}.

For the case of $SU(2)$-structure backgrounds with non-trivial NS three-form  the calibrations for odd cycles are given by (and here we assume no gauge field on the brane world-volumes )\footnote{Here we give calibrations for cycles defined on the internal space, an additional warp factor is required if the submanifold under consideration includes the space-time directions as in equation \ref{calizzz}.}
\beq
\Psi_{Cal~odd}=-8 h^{\frac{1}{4}}e^{-\frac{\Phi}{4}}\text{Im}(\Psi_{-})\wedge e^{B} \ , 
\eeq
while those for the even cycles by
\beq
\Psi_{Cal~even}=-8 h^{\frac{1}{4}}e^{-\frac{\Phi}{4}}\text{Im}(\Psi_{+})\wedge e^{B} \ , 
\eeq
where the pure spinors are given by equation (\ref{eq:SU2purespinor}) 
for $|ab|=e^{A}= \frac{e^{\frac{\Phi}{4}}}{h^{\frac{1}{4}}}$. 
Specifically this gives:
\begin{equation}
\begin{split}
&\mathcal{C}_1= - w_1 \ ,\\
&\mathcal{C}_2= - \text{Re} (\omega_2) \ ,\\
&\mathcal{C}_3= v_1\wedge j_2 - w_1\wedge B \ ,\\
&\mathcal{C}_4= -v_1\wedge w_1 \wedge\text{Im}(\omega_2)-B\wedge \text{Re}(\omega_2) \ ,\\
&\mathcal{C}_5= \frac{1}{2}w_1\wedge j_2\wedge j_2+v_1\wedge j_2 \wedge B-\frac{1}{2}w_1\wedge B\wedge B \ ,\\
&\mathcal{C}_6= -v_1\wedge w_1 \wedge\text{Im}(\omega_2)\wedge B -\frac{1}{2} \text{Re}(\omega_2)\wedge B\wedge B.
\end{split}
\end{equation}
A cycle in the internal space is supersymmetric if it satisfies the calibration condition 
\beq
\sqrt{g_{i} +B}d^i\xi =\mathcal{C}_i \ .  
\eeq

One can explicitly check that space-time filling D4, D6 and D8 branes wrapping the following cycles are indeed supersymmetric:
\be
\begin{aligned} 
	&\Sigma_{D4} =  (y^\mu , r)  \quad { \rm with} \quad  x_1= x_2 =0  \ , \\ 
	&\Sigma_{D6} =  (y^\mu, r, \psi , x_1  )   \quad { \rm with} \quad    x_1^2 + x_2^2 = const \ , \\
	&\Sigma_{D8} =   (y^\mu, r, \psi ,  \theta, \varphi  , x_1 )   \ .
\end{aligned}
\ee
We leave the task of finding other supersymmetric cycles for future work.

 \section{Discussion and Conclusions}

In this work we have clarified the action of non-Abelian T-duality in the context of backgrounds possessing $SU(3)\times SU(3)$ structure and ${\cal N}=1$ supersymmetry. 

We saw that rather generically the effect of performing a dualisation along an $SU(2)$ isometry group is to map an $SU(3)$-structure background to an $SU(2)$-structure background.  Such   geometries remain an interesting sector of compactifications which are much less well explored than their IIB $SU(3)$-structure cousins.  Our work then opens the door to constructing a rich class of such geometries.  Indeed although we have illustrated this work with the Klebanov-Witten geometry, everything we have said holds true for the wide variety of ${\cal N}=1$ backgrounds presented in \cite{Itsios:2013wd}  (details and extensions of this will appear in forthcoming work).  A particularly noteworthy direction is to consider the dualisation of more general toric Calabi-Yau geometries \cite{ISNTfuture}. 

One feature of the geometries presented above was that they possess {\em static} $SU(2)$-structure (that is the pure spinors are of type (2,1) everywhere).  An interesting question from the point of view of generalised complex geometry is whether backgrounds with a dynamic $SU(2)$-structure can be found using these techniques.  For this to be the case one would have to substantially change the relationship between the isometry group dualised and the initial complex structure.   

Establishing a clear dictionary between the geometries \cite{Itsios:2013wd}   discussed in this note  and a 
dual field theoretic description remains  the most pressing physical question. 
 In this note we showed how to readily add flavour branes to the picture and   this will provide further insight into any putative dual field theoretic description. 
 Indeed, the geometrical approach we started developing in this paper could extend
with interesting subtleties to the Klebanov-Strassler baryonic branch
solution (including the wrapped D5 system). This viewpoint will
 make clear the way to calculate some physical observables, like domain
walls and  other topological defects corresponding to branes wrapping
calibrated sub-manifolds.
On the other hand, it is likely that this geometric view might help
address important questions, like the periodicity of the new coordinates
 $x_1,x_2$, the existence of different cycles on which to integrate fluxes,
 a clear interpretation of the background in terms of color/flavor branes, etc.
All these points remain for future study. A long but somewhat clear path
 needs be travelled, to use the Maldacena Conjecture and define strongly coupled field theories
based on these backgrounds.

 \section*{Acknowledgements}

We wish to thank various colleagues for their feedback that helped 
improve this paper. Discussions with
Stefano Cremonesi, Akikazu Hashimoto, Chris Hull, Yolanda Lozano, 
Michela Petrini,  
Diego Rodriguez-Gomez, Alfonso Ramallo,
Kostas Sfetsos, 
Gary Shiu, 
Daniel Waldram, are greatefully acknowledged. 
A.B. acknowledges support from MECD FPU Grant No AP2009-3511 and projects FPA
2010-20807 and Consolider CPAN. A.B. would like to thank Swansea University Department
of Physics for hospitality.
The work of N.T Macpherson is supported by an STFC
studentship.The CP3-Origins Centre is partially funded by the Danish National Research Foundation under the research grant DNRF:90. The work of J.G.~was funded by the DOE Grant DE-FG02-95ER40896.

\appendix

\section{Appendix: Supergravity without Sources: conventions}

\label{appendixconventions}

We start by defining the Hodge star operator such that
\beq
\star\star F_{2n+1}=F_{2n+1},~~~\star\star F_{2n}=-F_{2n}.
\eeq
The action of type-IIB without sources is given in string frame by
\begin{equation}\label{Eq Action IIB}
\begin{split}
&S_{IIB}  = \int_{M_{10}}\!\!\!\! \sqrt{-g} \Bigg[e^{-2\Phi}
\left(R + 4 (\partial\Phi)^2 -\frac{H^2}{ 12}\right) -
\frac{1}{2}\left(F_1^2 + \frac{F_3^2}{ 3!} +\frac{1}{2}\frac{F_5^2}{ 5!}\right)
\Bigg] \\
&~~~~~~~~~~~~~~~~~~~~~~~~~ - \frac{1}{2}\left(C_4\wedge H \wedge dC_2\right).
\end{split}
\end{equation}
The fluxes can be conveniently defined as:
\begin{equation}
H=dB,~~~F_1=dC_0,~~~F_3=dC_2-H\wedge C_0,~~~F_5=dC_4- H\wedge C_2.
\end{equation}
These imply the following set of Bianchi identities 
(remind that there are no sources in the present section):
\beq
\begin{split}
&dH=0,~~~dF_1=0,~~~dF_3-H\wedge F_1=0,\\
& dF_5-H\wedge F_3=0.
\end{split}
\eeq
The dual fluxes, related by the expression $F_{2n+1}=(-)^{n}\star F_{9-2n}$, are defined as:
\beq
\star F_5=F_5,~~~F_7=dC_6-H\wedge C_4,~~~F_9=dC_8-H\wedge C_6 \ ,
\eeq
and the fluxes have the following equations of motion:
\beq
d\star F_1+H\wedge\star F_3=0,~~~~d\star F_3+H\wedge F_5=0.
\eeq
We can compactly express this in terms of the type IIB RR polyform as:
\beq
F_{IIB}=dC_{IIB}-H\wedge C_{IIB},
\eeq
where $C_{IIB}=C_0+C_2+C_4+C_6+C_8$. This has the combined Bianchi identity
\beq
dF_{IIB}=H\wedge F_{IIB}.
\eeq
The action of (massive) type IIA in string frame without sources is given by
\begin{equation}\label{Eq Action IIA}
\begin{split}
&S_{\rm Massive\ IIA}  = \int_{M_{10}}\!\!\!\! \sqrt{-g} \Bigg[e^{-2\Phi}
\left(R + 4 (\partial\Phi)^2 -\frac{H^2}{ 12}\right) -\frac{1}{2}\left(F_0^2 + \frac{F_2^2}{ 2} +\frac{F_4^2}{ 4!}\right)\Bigg] \\
&~~~~~~~~~~~~~~~~~~~~~~~~~ - \frac{1}{2}\left( dC_3 \wedge dC_3 \wedge B +  \frac{F_0}{ 3} dC_3 \wedge B^3 + \frac{F_0^2}{ 20} B^5\right) .
\end{split}
\end{equation}
The fluxes can be best expressed as
\beq
F_0= m,~~~F_2=dC_1+F_0 B,~~~F_4=dC_3 - H\wedge C_1 +\frac{F_0}{2}B\wedge B;
\eeq
where $m$ is a supergravity mass term. This leads to the following Bianchi identities
\beq
dF_0=0,~~~dF_2-F_0H=0,~~~dF_4-H\wedge F_2=0 \ .
\eeq
The dual fluxes, related by the expression $F_{2n}=(-)^{n}\star F_{10-2n}$, are defined as:
\beq
\begin{split}
&F_6=dC_5-H\wedge C_3+\frac{F_0}{3!}B^3,~~~F_8=dC_7-H\wedge C_5+\frac{F_0}{4!}F_0 B^4,\\
&F_{10}=dC_9-H\wedge C_7 +\frac{F_0}{5!}B^5.
\end{split}
\eeq
The flux equations of motion are:
\beq
d\star F_2+H\wedge\star F_4=0,~~~d\star F_4+H\wedge F_4=0.
\eeq
We can express this information in terms of the type IIA RR polyform as:
\beq
F_{IIA}=dC_{IIA}-H\wedge C_{IIA}+F_0 e^{B},
\eeq
where $C_{IIA}=C_1+C_3+C_5+C_7+C_9$. This has the combined Bianchi identity:
\beq
dF_{IIA}=H\wedge F_{IIA}.
\eeq
In appendix \ref{Sec: source conventions}, we will give expressions for the fluxes 
and their Bianchi identities in the presence of sources.

\section{Appendix: On $SU(2)$-Structures in six dimensions}\label{su2structuresxxx}

In this section, we give further details regarding the $SU(2)$-structure that are used through out the main body of this text. 
We sketch the derivation of the conditions that the $SU(2)$-structure must 
satisfy for $\mathcal{N}=1$ SUSY in type IIA. 
We will also use these to define potentials for the space-time 
filling RR-fluxes and the calibrations for space-time filling D4, D6 and D8 branes.
We assume a string frame metric of the form:
\beq
ds^2= e^{2A}dy_{1,3}+ ds_6^2
\eeq
with a dilaton $\Phi$ and a NS three form $H= dB$. We further assume that 
$\Phi(z), A(z)$ with $z$ any coordinate in $ds_6^2$.
Expanding out the $SU(2)$ pure spinors in \eqref{eq:SU2purespinor} gives:
\beq\label{eq: spinorexpansion}
\begin{split}
 & \Psi_+= \frac{|ab|}{8} \Big[\omega_2 -i \omega_2 \wedge v_1\wedge w_1 -\frac{1}{2}\omega_2 \wedge v_1\wedge w_1\wedge v_1\wedge w_1\Big]\ ,\\
 & \Psi_-= \frac{|ab|}{8} (1- i j_2 -\frac{1}{2} j_2\wedge j_2)\wedge (v_1 +i w_1)\ ,\\
 & \bar{\Psi}_-= \frac{|ab|}{8} \Big[v_1- i w_1 + j_2 \wedge (w_1+i v_1)-\frac{1}{2}j_2\wedge j_2 \wedge (v_1-i w_1)\Big]\ .
 \end{split}
\eeq
Supersymmetry requires that $|a| = |b|$, we define:
\beq
|ab| = |a|^2 = e^{A} \ .
\eeq
Plugging \eqref{eq: spinorexpansion} 
into \eqref{eq:purespinoreqs}, equating forms with 
equal number of legs and separating real and imaginary parts gives
\beq
	\begin{split}
 & d\Big[ e^{3A-\Phi}\omega_2  \Big]=0 \ ,\\
 & d\Big[ e^{3A-\Phi}\omega_2\wedge v_1\wedge w_1   \Big] 
+i  e^{3A-\Phi} H\wedge \omega_2=0 \ .
	\end{split}
\eeq
For two-forms,
\beq
	\begin{aligned}
		&d\Big[ e^{3A-\Phi}v_1  \Big] - e^{3A-\Phi} dA\wedge v_1 = 0 \ ,\\
		&d\Big[ e^{3A-\Phi}w_1 \Big] + e^{3A-\Phi} dA\wedge w_1= -e^{3A} \star_6 F_4 \ .
	\end{aligned}
\label{2-forms}
\eeq
For four-forms,
\beq
	\begin{aligned}
		&-d\Big[e^{3A-\Phi}j_2\wedge  w_1 \Big] - e^{3A-\Phi}H\wedge v_1+ e^{3A-\Phi}dA\wedge j_2 \wedge w_1= 0 \ ,\\
		&d\Big[e^{3A-\Phi}j_2\wedge v_1 \Big] - e^{3A-\Phi}H\wedge  w_1 + e^{3A-\Phi}dA\wedge j_2 \wedge v_1 = e^{3A} \star_6 F_2 \ ,
	\end{aligned}
\label{4-form}
\eeq
while for the six-form,
\beq
	\begin{aligned}
		&-\frac{1}{2}d\Big[e^{3A-\Phi}j_2\wedge j_2\wedge v_1 \Big] + e^{3A-\Phi}H \wedge j_2 \wedge w_1 +\frac{1}{2} e^{3A-\Phi} dA\wedge j_2\wedge j_2 \wedge v_1 = 0 \ ,\\
		& \frac{1}{2}d\Big[e^{3A-\Phi}j_2\wedge j_2\wedge w_1 \Big] + e^{3A-\Phi}H \wedge j_2 \wedge v_1 +\frac{1}{2} e^{3A-\Phi} dA\wedge j_2\wedge j_2 \wedge w_1=  e^{3A} \star_6 F_0 \ .
 	\end{aligned}
\label{6-form}
\eeq
Finally, we have for the zero-form
\beq
	\star_6 F_6 = 0
\eeq
where the fluxes $F_0$, $F_2$ and $F_4$ are understood to have legs in the six-dimensional internal space only. These equations can be further simplified as follows:
\begin{align}
		&d\Big[ e^{3A-\Phi}\omega_2  \Big]=0 \nonumber\\
		&\omega_2 \wedge \Big[ d\big(v_1\wedge w_1\big) + i H\Big] = 0 \nonumber\\
		&d\Big[ e^{2A-\Phi}v_1  \Big] = 0 \nonumber\\
		&d\Big[ e^{4A-\Phi}w_1 \Big] = -e^{4A} \star_6 F_4 \nonumber\\
		&d\Big[e^{2A-\Phi}j_2\wedge  w_1 \Big] + e^{2A-\Phi}H\wedge v_1= 0 \label{eq:SU2}\\
		&d\Big[e^{4A-\Phi}j_2\wedge v_1 \Big] - e^{4A-\Phi}H\wedge  w_1 = e^{4A} \star_6 F_2 \nonumber\\
		&\frac{1}{2}d\Big[e^{2A-\Phi}j_2\wedge j_2\wedge v_1 \Big] - e^{2A-\Phi}H \wedge j_2 \wedge w_1 = 0\nonumber\\
		&\frac{1}{2}d\Big[e^{4A-\Phi}j_2\wedge j_2\wedge w_1 \Big] + e^{4A-\Phi}H \wedge j_2 \wedge v_1= e^{4A} \star_6 F_0 \nonumber\\
		&\star_6 F_6 = 0.\nonumber
\end{align}
We clearly now have a definition of the Minkowski space-time filling RR-sector in terms of the $SU(2)$-structure:
\beq
\begin{split}
&F_6=d\Big[ e^{4A-\Phi}vol_4\wedge w_1 \Big]\\
&F_8=d\Big[e^{4A-\Phi}vol_4\wedge j_2\wedge v_1 \Big] - e^{4A-\Phi}H\wedge vol_4\wedge w_1\\
&F_{10}=-\frac{1}{2}d\Big[e^{4A-\Phi}vol_4\wedge j_2\wedge j_2\wedge w_1 \Big] + e^{4A-\Phi}H \wedge vol_4\wedge j_2 \wedge v_1,\\
\end{split}
\eeq
where the remaining fluxes can be obtained 
from the duality condition $F_{2n}=(-)^n\star F_{10-2n}$. 
With these equations it is possible to 
derive expressions for the potentials associated with these fluxes. They take the most compact 
form when the space-time filling part of the RR flux ployform 
is expressed as\footnote{We are assuming $B$ is defined only on the internal 
space so that $B^4=0$.}
\beq
F_{Mink}= d C_{Mink}-H\wedge C_{Mink} \ .
\eeq
We must have $-H\wedge C_3+\frac{1}{3!}F_0 B^3=0$ for $\mathcal{N}=1$ SUSY, otherwise the final line in 
equation \eqref{eq:SU2} cannot hold. This allows the derivation of canonical potentials in terms of the $SU(2)$-structure,
\be
\begin{split}
C_5 &= e^{4A-\Phi}vol_4\wedge w_1 \\ 
C_7 & =e^{4A-\Phi}vol_4\wedge j_2\wedge v_1\\
C_9 & =-\frac{1}{2}e^{4A-\Phi}vol_4\wedge j_2\wedge j_2 \wedge w_1 .  
\end{split} 
\ee
The calibration for type-IIA space-time filling D branes is defined as
\beq
\Psi_{cal} = - 8 e^{3A-\Phi} (\text{Im} \, \Psi_-) \wedge e^{B},
\eeq
expanding this out and extracting the terms with an equal number of legs gives:
\beq
\begin{split}
&\Psi^{(1)}_{cal}=-e^{4A-\Phi}w_1\\
&\Psi^{(3)}_{cal}=e^{4A-\Phi}\bigg(v_1\wedge j_2-w_1\wedge B\bigg)\\
&\Psi^{(5)}_{cal}=e^{4A-\Phi}\bigg(\frac{1}{2}w_1\wedge j_2\wedge j_2+v_1\wedge j_2\wedge B-\frac{1}{2}w_1\wedge B\wedge B\bigg) .
\end{split}
\label{calizzz}
\eeq
This makes it clear that an $SU(2)$-structure in six dimensions can 
potentially support Minkowski
space-time filling D4, D6 and D8 branes wrapping one, three, and five-cycles 
respectively.

\section{Appendix: Some Details of the Flavoured $SU(3)$ and $SU(2)$-structure solutions}\label{Sec: source conventions}

We will start analysing the case of the addition of flavours to the
Klebanov-Witten field theory \cite{Benini:2006hh}. This will be
explicitly dealt with using the language of
$SU(3)$-structures. Then, we will extend the analysis
to the background generated in section \ref{sectionkwflavored}. 
This will require the full
$SU(2)$-structure formalism, developed above.

We consider the addition of Minkowski space-time filling sources to an $SU(3)$-structure background in type-IIB.
The action of type-IIB in string frame is modified as:
\begin{equation}\label{Eq Action IIB plus sources}
S=S_{ IIB} +S_{DBI}+S_{WZ} .\\
\end{equation}
With pure spinors defined as in equation (\ref{Eq:SU3 pure spinor}) the calibration condition is given by:
\beq
\Psi_{Cal~IIB}= -8 e^{3A-\Phi} \bigg(\text{Im}\Psi_+\bigg) = e^{-\Phi}\bigg(\frac{e^{\Phi}}{h} \bigg)\bigg(1-\frac{1}{2} J\wedge J\bigg),
\eeq
which is compatible with source D3 and D7 branes.
We are assuming, as it is true for 
the Klebanov-Witten model  with massless flavours, that $H=0$.
The combined DBI action of such a system will be given by:
\beq
\begin{split}
&S_{DBI}= S^{D3}_{DBI}+S^{D7}_{DBI}, \\
&S^{D3}_{DBI}=-\int_{M_{10}} e^{-\Phi}\bigg(\frac{e^{\Phi}}{h} \bigg)vol_4\wedge \Xi_6, \\
&S^{D7}_{DBI}=\frac{1}{2}\int_{M_{10}} e^{-\Phi}\bigg(\frac{e^{\Phi}}{h} \bigg)vol_4\wedge J \wedge J\wedge\Xi_2.
\end{split}
\eeq
While the WZ terms will be given by:
\beq
\begin{split}
&S_{WZ}= S^{D3}_{WZ}+S^{D7}_{WZ}, \\
&S^{D3}_{WZ}=-\int_{M_{10}} C_4\wedge \Xi_6, \\
&S^{D7}_{WZ}=\int_{M_{10}} C_8\wedge \Xi_2.
\end{split}
\eeq
The fluxes, in the presence of sources -- for the case of $B=0$, should be defined as,
\begin{equation}
H=dB,~~~F_1=dC_0,~~~F_3=dC_2,~~~F_5=dC_4
\end{equation}
and the Bianchi identities are modified as follows:
\beq
\begin{split}
&dH=0,~~~dF_1=\Xi_2,~~~dF_3-H\wedge F_1=0 \ ,\\
& dF_5-H\wedge F_3=\Xi_6 \ ,
\end{split}
\eeq
where the $\Xi_i$'s that are non zero are determined by the specific source brane content.
The dual fluxes, related by the expression $F_{2n+1}=(-)^{n}\star F_{9-2n}$, are defined as:
\beq
\star F_5=F_5,~~~F_7=dC_6,~~~F_9=dC_8
\eeq
and the fluxes have the following equations of motion:
\beq
d\star F_1=0,~~~~d\star F_3=0 \ .
\eeq
For Klebanov-Witten with massless flavours we should set $\Xi_6=0$ and then the equation of motion of the dilaton and 
Einstein's equations can be shown to be satisfied also as in \cite{Nunez:2010sf}.

\subsection{Analysis of the generated background.}

In this work we  generated a flavoured type-IIA solution 
which supports an $SU(2)$-structure and non closed $B$. The action of (massive) type IIA in string frame, is now modified,
\begin{equation}\label{Eq Action IIA with sources}
S=S_{\rm Massive\ IIA} +S_{DBI}+S_{WZ}\\
\end{equation}
As shown around equation (\ref{calizzz}), an $SU(2)$-structure 
can in general support smeared source D4, D6 and D8 branes that extend 
in the Minkowski directions. The combined DBI and WZ actions of this system are given by: 
\beq
	\begin{aligned}
		S_{DBI} &= S^{D8}_{DBI}+S^{D6}_{DBI}+S^{D4}_{DBI} \\
		S^{D4}_{DBI} &= \int_{M_{10}}\!\!\!\! e^{-\hat{\Phi}}\left(\frac{e^{\Phi}}{h} vol_4\right) \wedge w_1\wedge\Xi_5,\\
		S^{D6}_{DBI} &= -\int_{M_{10}}\!\!\!\! e^{-\hat{\Phi}}\left(\frac{e^{\Phi}}{h} vol_4\right) \wedge\bigg(v_1\wedge j_2-w_1\wedge B\bigg)\wedge\Xi_3, \label{BIxx}\\
		S^{D8}_{DBI} &= -\int_{M_{10}}\!\!\!\! e^{-\hat{\Phi}} \left(\frac{e^{\Phi}}{h}vol_4\right)\wedge\left(\frac{1}{2}w_1\wedge j_2\wedge j_2+v_1\wedge j_2\wedge B-\frac{1}{2}w_1\wedge B\wedge B\right)\wedge\Xi_1\ ,
	\end{aligned}
\eeq
and
\begin{equation}
\begin{split}
S_{WZ}&=S^{D8}_{WZ}+S^{D6}_{WZ}+S^{D4}_{WZ},\\
S^{D4}_{WZ}&=-\int_{M_{10}}\!\!C_5\wedge\Xi_5, \\\
S^{D6}_{WZ}&=\int_{M_{10}}\!\!\bigg(C_7-B\wedge C_5\bigg)\wedge\Xi_3, \\\
S^{D8}_{WZ}&=-\int_{M_{10}}\!\!\left(C_9-B\wedge C_7+\frac{1}{2}B
\wedge B\wedge C_5\right)\wedge\Xi_1.\\
\label{WZxx}
\end{split}
\end{equation}
In the presence of such sources we should define the RR-potentials as:
\beq
F_0,~~~F_2=dC_1+F_0 B,~~~F_4=dC_3+B\wedge dC_1+\frac{F_0}{2}B\wedge B;
\eeq
this ensures that we have no ill-defined potential terms appearing explicitly. 
We note that source D8 branes imply that $F_0$ will 
no longer be quantised. In general the Bianchi identities are given by
\beq
\begin{split}
&dF_0=\Xi_1,~~~dF_2-F_0H=\Xi_3+B\wedge \Xi_1;\\
&dF_4-H\wedge F_2=\Xi_5+B\wedge\Xi_3+\frac{1}{2}B\wedge B\wedge\Xi_1 .
\end{split}
\label{defes}
\eeq
The dual fluxes, related by the expression $F_{2n}=(-)^{n}\star F_{10-2n}$, are defined as:
\beq
\begin{split}
&F_6=dC_5,~~~F_8=dC_7-H\wedge C_5,\\
&F_{10}=dC_9-H\wedge C_7.
\end{split}
\eeq
Here, we did not write 
the terms that are zero due to the $SU(2)$ SUSY conditions in six dimensions. 
The flux equations of motion for the RR sector are given by:
\beq
d\star F_2+H\wedge\star F_4=0,~~~d\star F_4+H\wedge F_4=0,
\eeq
while for the  NS sector we find:
\begin{equation}
\begin{split}
d\left(e^{-2\hat{\Phi}}\star H\right)=&F_0 \star F_2+  F_2\wedge \star F_4  
+\frac{1}{2}F_4\wedge F_4 -\\  
&\frac{e^{\Phi-\hat{\Phi}}}{h}\bigg[vol_4\wedge\left(
w_1\wedge B-v_1\wedge j_2\right)\wedge\Xi_1+
vol_4\wedge w_1\wedge\Xi_3\bigg] \ .
\end{split}
\end{equation}
A careful calculation shows that the potentials 
do not enter into this equation explicitly \cite{Koerber:2007hd}.
We can express the variation of the dilaton as an integral for compactness,
\begin{equation}\label{Eq.DBI check}
S_{DBI}=-\int8e^{-2\hat{\Phi}}(d\star d\hat{\Phi}+\star\frac{R}{4}-d\hat{\Phi}\wedge\star d\hat{\Phi}-\frac{1}{8}H\wedge \star H) \ .\\
\end{equation}
It is useful at this stage to introduce the following notation,
\begin{equation}
	\omega_{(p)}\lrcorner \lambda_{(p)} = \frac{1}{p!} \omega^{\mu_1 ... \mu_p} \lambda_{\mu_1 ... \mu_p}
\end{equation}
where the following identity is helpful,
\begin{equation}
	\int \omega_{(p)} \wedge \lambda_{(10-p)} = - \int \sqrt{-g} \lambda \lrcorner  (\star\omega).
\end{equation}
Then Einstein's equations can be expressed in a gauge-invariant fashion as:
\begin{equation}
\begin{split}
R_{\mu\nu}=&-2D_{\mu}D_{\nu}\hat{\Phi}+\frac{1}{4}H^2_{\mu\nu}+e^{2\hat{\Phi}}\bigg[\frac{1}{2} (F_2^2)_{\mu\nu}+\frac{1}{12}(F_4^2)_{\mu\nu}-\frac{1}{4}g_{\mu\nu}(F_0^2+\frac{1}{2}F_2^2+\frac{1}{4!}F_4^2)\bigg]+\\
&\frac{e^{\Phi+\hat{\Phi}}}{h}\bigg[\frac{1}{48}(\Xi_5+\Xi_3\wedge B+\frac{1}{2}B \wedge B \wedge \Xi_1)_{\mu \alpha_1...\alpha_4}\star (vol_4\wedge w_1)_{\nu}^{\alpha_1...\alpha_4}-\\
&~~~~~~~~~\frac{1}{4}(\Xi_3+B\wedge \Xi_1)_{\mu\alpha_1\alpha_2}\star (vol_4\wedge v_1\wedge j_2)_{\nu}^{\alpha_1\alpha_2} - \frac{1}{4}\Xi_1\!~_{\mu}\star(vol_4\wedge w_1\wedge j_2 \wedge j_2)_{\nu}\\
&~~~~~~~~-\frac{1}{4}g_{\mu\nu}\bigg((\Xi_5+\Xi_3\wedge B+\frac{1}{2}B \wedge B \wedge \Xi_1)\lrcorner\star (vol_4\wedge w_1)-\\
&~~~~~~~~~(\Xi_3+B\wedge \Xi_1)\lrcorner\star (vol_4\wedge v_1\wedge j_2) - \frac{1}{2}\Xi_1\lrcorner\star(vol_4\wedge w_1\wedge j_2 \wedge j_2)\bigg) \bigg].
\end{split}
\label{einsteinsvvv}
\end{equation}
The equations (\ref{defes})-(\ref{einsteinsvvv}) are solved by the system in section \ref{sectionkwflavored} after the BPS
equations (\ref{eq:BPS}) are imposed.

\providecommand{\href}[2]{#2}\begingroup\raggedright\endgroup


\begin{thebibliography}{99}



\bibitem{delaossa:1992vc}
X.C. de~la Ossa and F.~Quevedo, {\it Duality symmetries from non abelian isometries in string theory},  Nucl. Phys. {\bf B403} (1993) 377, \href{http://arxiv.org/abs/hep-th/9210021}{{\tt hep-th/9210021}}.



\bibitem{Sfetsos:2010uq}
  K.~Sfetsos and D.~C.~Thompson,
  {\em On non-Abelian T-dual geometries with Ramond fluxes},
  Nucl.\ Phys.\ B {\bf 846} (2011) 21
  \href{http://arxiv.org/abs/1012.1320}{{\tt arXiv:1012.1320}}.

 

\bibitem{Lozano:2011kb}
  Y.~Lozano, E.~.O Colgain, K.~Sfetsos and D.~C.~Thompson,
  {\em Non-Abelian T-duality, Ramond Fields and Coset Geometries},
    JHEP {\bf 1106} (2011) 106
    \href{http://arxiv.org/abs/1104.5196}{{\tt arXiv:1104.5196}}.

 

\bibitem{Itsios:2012dc}
  G.~Itsios, Y.~Lozano, E.~.O Colgain and K.~Sfetsos,
{\em Non-Abelian T-duality and consistent truncations in type-II supergravity},
  JHEP {\bf 1208} (2012) 132
      \href{http://arxiv.org/abs/1205.2274}{{\tt arXiv:1205.2274}}.

 

\bibitem{Lozano:2012au}
  Y.~Lozano, E.~O Colgain, D.~Rodriguez-Gomez and K.~Sfetsos,
  {\em New Supersymmetric $AdS_6$ via T-duality},
       \href{http://arxiv.org/abs/1212.1043}{{\tt arXiv:1212.1043}}.
  
\bibitem{Jeong:2013jfc}
  J.~Jeong, \"Oz.~\"ur~Kelekci and E.~\'O~Colg\'ain,
 {\em An alternative IIB embedding of F(4) gauged supergravity},
  JHEP {\bf 1305} (2013) 079
         \href{http://arxiv.org/abs/1302.2105}{{\tt arXiv:1302.2105}}.


\bibitem{Itsios:2012zv}
  G.~Itsios, C.~Nunez, K.~Sfetsos and D.~C.~Thompson,
  {\em On non-Abelian T-Duality and new N=1 backgrounds},
   Phys.\ Lett.\ B {\bf 721} (2013) 342
         \href{http://arxiv.org/abs/1212.4840}{{\tt arXiv:1212.4840}}.

  

\bibitem{Itsios:2013wd}
  G.~Itsios, C.~Nunez, K.~Sfetsos and D.~C.~Thompson,
 {\em Non-Abelian T-duality and the AdS/CFT correspondence:new N=1 backgrounds},
    Nucl.\ Phys.\ B {\bf 873} (2013) 1
     \href{http://arxiv.org/abs/1301.6755}{{\tt arXiv:1301.6755}}.


\bibitem{Bah:2012dg}
  I.~Bah, C.~Beem, N.~Bobev and B.~Wecht,
{\em Four-Dimensional SCFTs from M5-Branes},
  JHEP {\bf 1206} (2012) 005
       \href{http://arxiv.org/abs/1203.0303}{{\tt arXiv:1203.0303}}. 




\bibitem{Grana:2004bg}
  M.~Grana, R.~Minasian, M.~Petrini and A.~Tomasiello,
  {\em Supersymmetric backgrounds from generalized Calabi-Yau manifolds},
  JHEP {\bf 0408} (2004) 046
Phys.\ Lett.\ B {\bf 355} \href{http://arxiv.org/abs/hep-th/0406137}{{\tt hep-th/0406137}}.


\bibitem{Grana:2005sn} 
  M.~Grana, R.~Minasian, M.~Petrini and A.~Tomasiello,
  {\em Generalized structures of N=1 vacua},
  JHEP {\bf 0511}, 020 (2005)
   \href{http://arxiv.org/abs/hep-th/0505212}{{\tt hep-th/0505212}}.

  
\bibitem{Strominger:1996it}
  A.~Strominger, S.~-T.~Yau and E.~Zaslow,
{\em Mirror symmetry is T duality},
  Nucl.\ Phys.\ B {\bf 479} (1996) 243
   \href{http://arxiv.org/abs/hep-th/9606040}{{\tt hep-th/9606040}}.
 

\bibitem{Gurrieri:2002wz}
  S.~Gurrieri, J.~Louis, A.~Micu and D.~Waldram,
{\em Mirror symmetry in generalized Calabi-Yau compactifications},
  Nucl.\ Phys.\ B {\bf 654} (2003) 61
     \href{http://arxiv.org/abs/hep-th/0211102}{{\tt hep-th/0211102}}. 

\bibitem{Fidanza:2003zi}
  S.~Fidanza, R.~Minasian and A.~Tomasiello,
  {\em Mirror symmetric SU(3) structure manifolds with NS fluxes},
  Commun.\ Math.\ Phys.\  {\bf 254} (2005) 401
     \href{http://arxiv.org/abs/hep-th/0311122}{{\tt hep-th/0311122}}.


\bibitem{Grana:2006hr}
  M.~Grana, J.~Louis and D.~Waldram,
 {\em SU(3) x SU(3) compactification and mirror duals of magnetic fluxes},
  JHEP {\bf 0704} (2007) 101
     \href{http://arxiv.org/abs/hep-th/0612237}{{\tt hep-th/0612237}}. 

\bibitem{Grana:2008yw}
  M.~Grana, R.~Minasian, M.~Petrini and D.~Waldram,
  {\em T-duality, Generalized Geometry and Non-Geometric Backgrounds},
  JHEP {\bf 0904} (2009) 075 
           \href{http://arxiv.org/abs/0807.4527}{{\tt arXiv:0807.4527}}.


\bibitem{Erdmenger:2007cm} 
  J.~Erdmenger, N.~Evans, I.~Kirsch and E.~Threlfall,
{\em Mesons in Gauge/Gravity Duals - A Review},
  Eur.\ Phys.\ J.\ A {\bf 35}, 81 (2008)
             \href{http://arxiv.org/abs/0711.4467}{{\tt arXiv:0711.4467}}.



\bibitem{Nunez:2010sf} 
  C.~Nunez, A.~Paredes and A.~V.~Ramallo,
  {\em Unquenched Flavor in the Gauge/Gravity Correspondence},
  Adv.\ High Energy Phys.\  {\bf 2010}, 196714 (2010)
               \href{http://arxiv.org/abs/1002.1088}{{\tt arXiv:1002.1088}}.

  F.~Bigazzi, A.~L.~Cotrone, J.~Mas, D.~Mayerson and J.~Tarrio,
{\em Holographic Duals of Quark Gluon Plasmas with Unquenched Flavors},
  Commun.\ Theor.\ Phys.\  {\bf 57}, 364 (2012)
                 \href{http://arxiv.org/abs/1110.1744}{{\tt arXiv:1110.1744}}.










\bibitem{Casero:2006pt} 
  R.~Casero, C.~Nunez and A.~Paredes,
 {\em Towards the string dual of N=1 SQCD-like theories},
  Phys.\ Rev.\ D {\bf 73}, 086005 (2006)
       \href{http://arxiv.org/abs/hep-th/0602027}{{\tt hep-th/0602027}}.

  F.~Benini, F.~Canoura, S.~Cremonesi, C.~Nunez and A.~V.~Ramallo,
  {\em Backreacting flavors in the Klebanov-Strassler background},
  JHEP {\bf 0709}, 109 (2007)
                   \href{http://arxiv.org/abs/0706.1238}{{\tt arXiv:0706.1238}}.

  E.~Conde, J.~Gaillard, C.~Nunez, M.~Piai and A.~V.~Ramallo,
{\em A Tale of Two Cascades: Higgsing and Seiberg-Duality Cascades from type IIB String Theory},
  JHEP {\bf 1202}, 145 (2012)
                     \href{http://arxiv.org/abs/1112.3350}{{\tt arXiv:1112.3350}}.

  N.~T.~Macpherson,
  {\em The Holographic Dual of 2+1 Dimensional QFTs with N=1 SUSY and Massive Fundamental Flavours},
  JHEP {\bf 1206}, 136 (2012)
              \href{http://arxiv.org/abs/1204.4222}{{\tt arXiv:1204.4222}}. 

  N.~T.~Macpherson,
{\em SuGra on G2 Structure Backgrounds that Asymptote to AdS4 and Holographic Duals of Confining 2+1d Gauge Theories with N=1 SUSY},
  JHEP {\bf 1304} (2013) 076
                \href{http://arxiv.org/abs/1301.5178}{{\tt arXiv:1301.5178}}. 





\bibitem{Martucci:2005ht} 
  L.~Martucci and P.~Smyth,
{\em Supersymmetric D-branes and calibrations on general N=1 backgrounds},
  JHEP {\bf 0511}, 048 (2005)
         \href{http://arxiv.org/abs/hep-th/0507099}{{\tt hep-th/0507099}}.





\bibitem{Koerber:2007hd} 
  P.~Koerber and D.~Tsimpis,
  {\em Supersymmetric sources, integrability and generalized-structure compactifications},
  JHEP {\bf 0708}, 082 (2007)
                  \href{http://arxiv.org/abs/0706.1244}{{\tt arXiv:0706.1244}}.  



\bibitem{Gaillard:2008wt} 
  J.~Gaillard and J.~Schmude,
 {\em On the geometry of string duals with backreacting flavors},
  JHEP {\bf 0901}, 079 (2009)
                    \href{http://arxiv.org/abs/0811.3646}{{\tt arXiv:0811.3646}}.  



\bibitem{Gaillard:2010qg} 
  J.~Gaillard, D.~Martelli, C.~Nunez and I.~Papadimitriou,
 {\em The warped, resolved, deformed conifold gets flavoured},
  Nucl.\ Phys.\ B {\bf 843}, 1 (2011)
          \href{http://arxiv.org/abs/1004.4638}{{\tt arXiv:1004.4638}}.

\bibitem{Mariotti:2007ym} 
  A.~Mariotti,
{\em Supersymmetric D-branes on SU(2) structure manifolds},
  JHEP {\bf 0709}, 123 (2007)
            \href{http://arxiv.org/abs/0705.2563}{{\tt arXiv:0705.2563}}.



\bibitem{Curtright:1994be}
  T.~Curtright and C.~K.~Zachos,
{\em Currents, charges, and canonical structure of pseudodual chiral models},
  Phys.\ Rev.\ D {\bf 49} (1994) 5408
  \href{http://arxiv.org/abs/hep-th/9401006}{{\tt hep-th/9401006}}. 


\bibitem{Berkovits:2008ic}
  N.~Berkovits and J.~Maldacena,
{\em Fermionic T-Duality, Dual Superconformal Symmetry, and the Amplitude/Wilson
 Loop Connection},  
JHEP {\bf 0809} (2008) 062 
   \href{http://arxiv.org/abs/0807.3196}{{\tt arXiv:0807.3196}}. 

\bibitem{Alday:2007hr}
  L.~F.~Alday and J.~M.~Maldacena,
 {\em Gluon scattering amplitudes at strong coupling},  
JHEP {\bf 0706} (2007) 064 
   \href{http://arxiv.org/abs/0705.0303}{{\tt arXiv:0705.0303}}. 



\bibitem{Lozano:1995jx}
  Y.~Lozano,
  {\em NonAbelian duality and canonical transformations},  
Phys.\ Lett.\ B {\bf 355} \href{http://arxiv.org/abs/hep-th/9503045}{{\tt hep-th/9503045}}.


\bibitem{Benichou:2008it} 
  R.~Benichou, G.~Policastro and J.~Troost,
 {\em T-duality in Ramond-Ramond backgrounds},
  Phys.\ Lett.\ B {\bf 661}, 192 (2008)
     \href{http://arxiv.org/abs/0801.1785}{{\tt arXiv:0801.1785}}. 



\bibitem{Sfetsos:2010xa} 
  K.~Sfetsos, K.~Siampos and D.~C.~Thompson,
  {\em Canonical pure spinor (Fermionic) T-duality},
  Class.\ Quant.\ Grav.\  {\bf 28}, 055010 (2011)
       \href{http://arxiv.org/abs/1007.5142}{{\tt arXiv:1007.5142}}. 


\bibitem{Hassan:1999bv}
  S.~F.~Hassan,
{\em T duality, space-time spinors and RR fields in curved backgrounds},
  Nucl.\ Phys.\ B {\bf 568} (2000) 145
  \href{http://arxiv.org/abs/hep-th/9907152}{{\tt hep-th/9907152}}.\\
  S.~F.~Hassan,
{\em SO(d,d) transformations of Ramond-Ramond fields and space-time spinors},
  Nucl.\ Phys.\ B {\bf 583} (2000) 431
\href{http://arxiv.org/abs/hep-th/9912236}{{\tt hep-th/9912236}}.

\bibitem{Kosmann}
Y.~Kosmann, {\it {A note on Lie-Lorentz derivatives}},   Annali di Mat.
  Pura Appl. {\bf (IV) 91} (1972) 317. \\
J.~M. Figueroa-O'Farrill, {\it {On the supersymmetries of anti de Sitter
  vacua}},  Class. Quant. Grav. {\bf 16} (1999) 2043,
  \href{http://xxx.lanl.gov/abs/hep-th/9902066}{{\tt hep-th/9902066}}.
 \\
 T.~Ortin, {\it {A note on Lie-Lorentz derivatives}},  Class. Quant. Grav.
  {\bf 19} (2002) L143,
  \href{http://xxx.lanl.gov/abs/hep-th/0206159}{{\tt hep-th/0206159}}.


 \bibitem{Benini:2006hh} 
  F.~Benini, F.~Canoura, S.~Cremonesi, C.~Nunez and A.~V.~Ramallo
{\em Unquenched flavors in the Klebanov-Witten model},
  JHEP {\bf 0702}, 090 (2007)
  \href{http://arxiv.org/abs/hep-th/0612118}{{\tt hep-th/0612118}}.




\bibitem{Klebanov:1998hh}
  I.~R.~Klebanov and E.~Witten,
{\em Superconformal field theory on three-branes at a Calabi-Yau singularity},
  Nucl.\ Phys.\ B {\bf 536} (1998) 19
    \href{http://arxiv.org/abs/hep-th/9807080}{{\tt hep-th/9807080}}.





\bibitem{Marolf:2000cb}
  D.~Marolf,
{\em Chern-Simons terms and the three notions of charge},
    \href{http://arxiv.org/abs/hep-th/0006117}{{\tt hep-th/0006117}}.

 \bibitem{ISNTfuture}
  G.~Itsios, C.~Nunez, K.~Sfetsos and D.~C.~Thompson, forthcoming.


\end{thebibliography}
\end{document}